\documentclass[a4paper,11pt]{article}
\pdfoutput=1 

\usepackage{jcappub} 

\usepackage[T1]{fontenc} 
\usepackage[font={small}]{caption}
\usepackage{subcaption}

\title{\boldmath Hemispherical Power Asymmetry in intensity and polarization for Planck PR4 data}

\author[a, b]{C. Gimeno-Amo,}
\author[a]{R. B. Barreiro,}
\author[a]{E. Martínez-González}
\author[c]{and A. Marcos-Caballero}

\affiliation [a]{Instituto de Física de Cantabria, CSIC-Universidad de Cantabria,\newline Avda. de los Castros s/n, 39005 Santander, Spain.}
\affiliation[b]{Departamento de Física Moderna, Universidad de Cantabria,\newline Avda. de los Castros s/n, 39005 Santander, Spain.}
\affiliation[c]{Department of Theoretical Physics, University of the Basque Country UPV/EHU,\newline Barrio Sarriena s/n, Leioa, Vizcaya, Spain}

\emailAdd{gimenoc@ifca.unican.es}
\emailAdd{barreiro@ifca.unican.es}
\emailAdd{martinez@ifca.unican.es}
\emailAdd{marcos@ifca.unican.es}

\abstract{One of the foundations of the Standard Model of Cosmology is statistical isotropy, which can be tested, among other probes, through the study of the Cosmic Microwave Background (CMB). However, a hemispherical power asymmetry on large scales has been reported for WMAP and Planck data by different works. 
The statistical significance is above 3$\sigma$ for temperature, suggesting a directional dependence of the local power spectrum, and thus a feature beyond the $\Lambda$CDM model. With the third release of the Planck data (PR3), a new analysis was performed including the E-mode polarization maps, finding an asymmetry at a modest level of significance. In this work, we perform an asymmetry analysis in intensity and polarization maps for the latest Planck processing pipeline (PR4). We obtain similar results to those obtained with PR3, with a slightly lower significance (2.8\% for the Sevem method)
for the amplitude of the E-mode local variance dipole as well as a significant variability with the considered mask.
In addition, a hint of a possible T-E alignment between the asymmetry axes is found at the level of $\sim$ 5\%. For the analysis, we have implemented an alternative inpainting approach in order to get an accurate reconstruction of the E-modes. More sensitive all-sky CMB polarization data, such as those expected from the future LiteBIRD experiment, are needed to reach a more robust conclusion on the possible existence of deviations from statistical isotropy in the form of a hemispherical power asymmetry.}

\begin{document}
\maketitle
\flushbottom

\section{Introduction}

In the standard $\Lambda$CDM model the statistical isotropy of the universe is considered as one of the main foundations. This idea is supported by the simplest slow-roll inflationary scenarios, where the primordial fluctuations, the seeds for late universe structures, are a single realization of a statistically isotropic Gaussian random field \cite{2020A&A...641A..10P}. These properties can be accurately tested trough the Cosmic Microwave Background (CMB) anisotropies.

Observations of the CMB by the Planck satellite are consistent with the Gaussian, almost scale-invariant, and adiabatic nature of the primordial perturbations predicted by the slow-roll inflation, providing a very accurate confirmation of the base $\Lambda CDM$ model \cite{Planck:2018nkj, 2020A&A...641A...5P, 2020A&A...641A...6P, Planck:2019evm}. However, some unexpected features on large angular scales have been found in the temperature fluctuation field. These anomalies were already observed soon after the first release of the NASA \textit{Wilkinson Microwave Anisotropy Probe} (WMAP) satellite data \cite{2003ApJS..148....1B, 2007ApJS..170..288H, 2009ApJS..180..225H, 2011ApJS..192...17B, 2013ApJS..208...19H}, and they show tensions with the standard model. Some of them suggest a potential violation of the cosmological principle, while others are related to non-Gaussian imprints. Among these features are the power asymmetry \cite{2004ApJ...605...14E, 2004MNRAS.354..641H}, quadrupole-octopole alignment\cite{2004PhRvD..69f3516D}, the cold spot \cite{2004ApJ...609...22V, 2010AdAst2010E..77V}, the anomalously low variance \citep{Monteserin08} or the lack of power at large scales \cite{2006A&A...454..409B}. In \cite{2023CQGra..40i4001K} a summary of some of the anomalies observed in the CMB can be found.

In the present work, we are interested in the "hemispherical power asymmetry" (HPA). This anomalous feature has been found with different methods at the level of 3$\sigma$-3.5$\sigma$ depending on the estimator and the angular scales used for the analysis \cite{2007ApJ...660L..81E, 2009ApJ...704.1448H, 2009ApJ...699..985H, 2014A&A...571A..23P, Akrami:2014eta, 2015MNRAS.446.4232A, 2016A&A...594A..16P, Planck:2019evm}. The asymmetry persists in Planck data, ruling out systematics as a plausible explanation, as it has been observed at a similar level by two independent missions.

The HPA has been modeled as a dipole modulation of the primordial fluctuations \cite{2005PhRvD..72j3002G, 2007ApJ...656..636G} as
\begin{equation}\label{DM}
    \Delta{\tilde{T}}(\hat{n}) = (1+A_{DM}\hat{n}\cdot\hat{p})\Delta{T}(\hat{n})
\end{equation}
where $\Delta{\tilde{T}}$ and $\Delta{T}$ are the modulated and isotropic CMB fluctuation fields, respectively, and $A_{DM}$ and $\hat{p}$ the amplitude and direction of the modulation.

Some analyses have 
performed a direct fitting
to this model \cite{2009ApJ...699..985H, 2007ApJ...660L..81E, 2014A&A...571A..23P, 2017PhRvD..95f3011Z}. They found amplitudes close to 7\% 
at a confidence level
around 3$\sigma$. Equivalent results have been obtained following methods working in 
spherical harmonics space \cite{2015PhRvD..91b3515R, 2009PhRvD..80f3004H, 2016JCAP...01..046G, 2016A&A...594A..16P}. These estimators exploit the expected correlations between $\ell$ and $\ell\pm 1$ induced by the possible dipolar modulation. The HPA has also been formulated in terms of Bipolar spherical harmonics (BipoSH) \cite{2016A&A...594A..16P, 2023arXiv230104539D} technique, which works in the harmonic space and has the capability of tracing 
statistical isotropy violations.

It has also been observed that the dipolar modulation is not scale-independent in the sense that the amplitude is much lower for small angular scales and apparently vanishes for multipoles above six hundred \cite{2009ApJ...704.1448H, Akrami:2014eta, Marcos-Caballero:2019jqj}. In particular, in most analyses, the statistical significance decreases if the full range of scales is taken into account.
In any case, the dipolar power asymmetry remains an intriguing anomaly present on super-Hubble scales at the recombination time, suggesting a possible physical origin related to the inflationary mechanism \cite{2016JCAP...01..046G, 2016A&A...594A..16P}.

Although most of the analyses have assummed a dipolar modulation as the fitting model, it is important to remark that this is not necessarily the underlying model of the HPA. There have been some attempts to propose physical models that could explain the origin of the HPA and reconcile observations with the standard cosmological model \cite{2008PhRvD..78l3520E, 2009PhRvD..80b3526D, 2010PhRvD..81h3501C, 2011JHEP...02..061K, 2013JCAP...08..007L, 2016MNRAS.460.1577K, 2017PhRvD..95f3011Z, 2022arXiv220903928S}. Some of them suggest a superhorizon perturbation as a remnant of a pre-inflationary phase where inflation is given in a multi-field scenario, including curvaton-type models \cite{2008PhRvD..78l3520E}. This type of models predicts non-Gaussian imprints in the CMB maps, which can be used to constrain these scenarios. However, according to the latest Planck inflationary constraints, 
they are not statistically favored compared to $\Lambda$CDM \cite{2020A&A...641A..10P}. A non-cosmological explanation has been recently proposed related to a new foreground associated to nearby galaxies \cite{Hansen:2023gra}.

In this paper, we perform an analysis similar to the one carried out by the Planck collaboration in \cite{Planck:2019evm} using the latest Planck Release 4 (PR4) data set, also known as NPIPE, in both intensity and polarization E-mode maps. In particular, we use a pixel space \emph{local-variance} method introduced in \cite{Akrami:2014eta}. We study the presence of the HPA in the intensity clean CMB maps 
obtained from PR4 using the Sevem \cite{2012MNRAS.420.2162F} and Commander \cite{2008ApJ...676...10E} componet separation methods. In polarization only Sevem data is used.

Exploring the asymmetry in polarization data is very important to answer the question of whether the HPA is related to new physics or a statistical fluke, as the large-scale intensity data on its own is not able to claim a strong positive detection and is already cosmic-variance limited. Furthermore, as discussed in \cite{2020MNRAS.492.3994G}, different proposed models for the HPA in intensity will also predict a similar asymmetry in CMB polarization data.
There have been some attempts to estimate the significance level of the HPA in the Planck E-mode map \cite{2017arXiv171000580A, 2020MNRAS.492.3994G, Planck:2019evm}. Using the PR3 data set, Planck collaboration found a large dispersion in the p-value between the four component separation methods, reflecting the presence of different amounts of residuals in the corresponding maps. Actually, Sevem and Commander returned a p-value below 1\%, while NILC and SMICA showed a p-value closer to 5\%-6\%.

The structure of the paper is as follows. In section \ref{DatSim} we describe the data and simulations used in the analysis. The methodology is outlined in section \ref{Met}, where we describe in detail the inpainting approach, the local-variance estimator, and the procedure followed for generating the masks. The main results are shown in Section \ref{Res}. Finally, a summary of the main conclusions from the analysis is given in section \ref{Con}.

\section{Data and Simulations}\label{DatSim}

\begin{figure}[t!]
\centering
  \includegraphics[scale = 0.43]{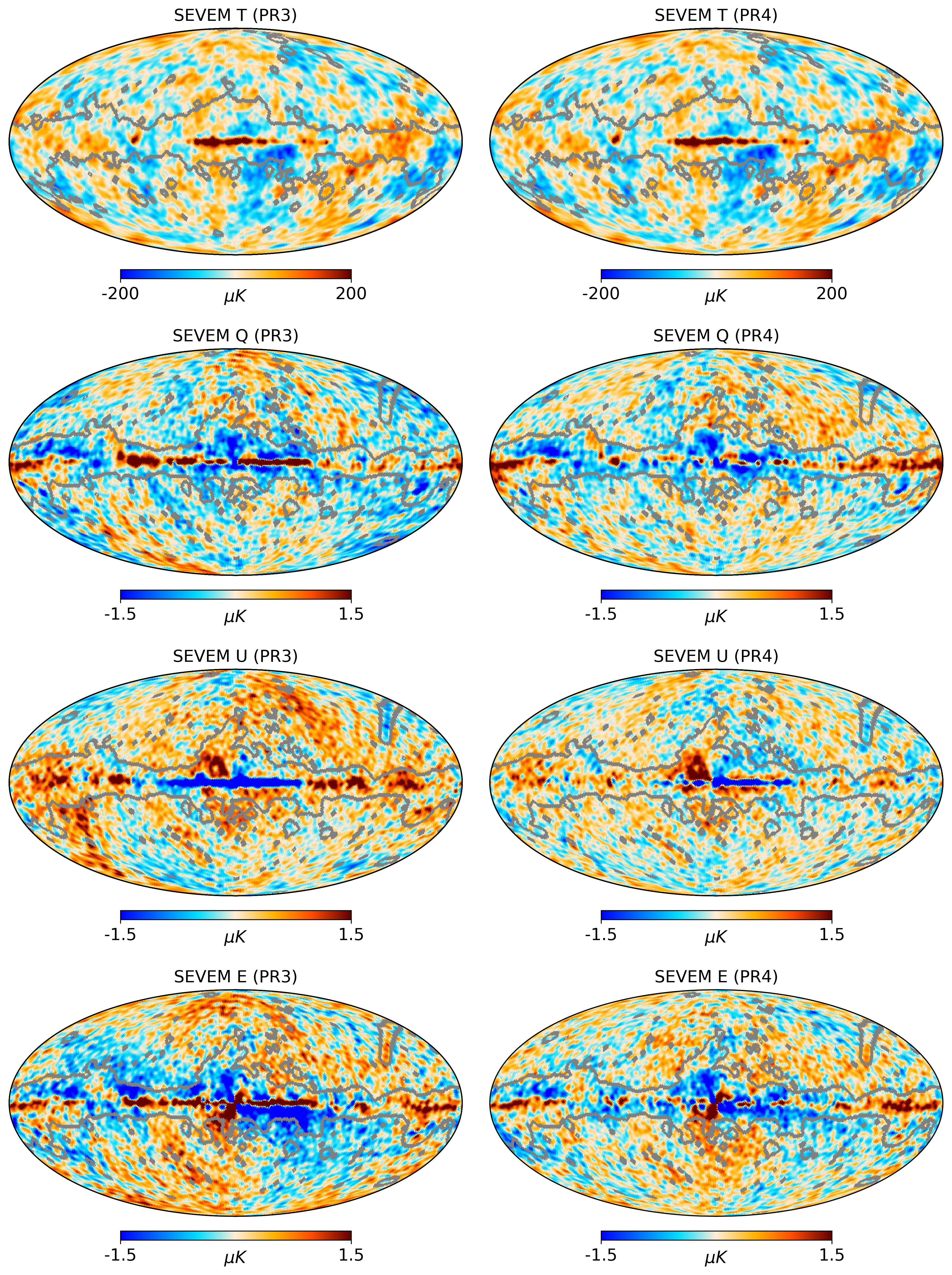}
  \caption{Sevem component-separated CMB maps at 160$'$ resolution. The first column shows T, Q, U, and E-mode maps for the Planck 2018 pipeline (PR3), while the second column shows the same maps for NPIPE (PR4). E-mode maps are derived from the full-sky Stokes Q and U maps. The grey lines indicate either the temperature or the
  polarization Planck component separation confidence mask  \cite{Planck:2018yye}.}
  \label{fig:Maps}
\end{figure}

In the present analysis we have used the Planck data\footnote{Planck space telescope, operated by the ESA, had two instruments, the Low Frequency Instrument (LFI) and the High Frequency Instrument (HFI), with the goal of measuring the total intensity and polarization of CMB photons in a wide frequency range from 30 to 857 GHz (30 to 353 for polarization). This range was 
covered by 9 different frequencies, with the three lowest frequencies measured by the LFI (30, 44 and 70 GHz) and the rest by the HFI (100, 143, 217, 353, 545 and 857 GHz).} from the latest two releases, the Planck 2018 full-mission release (PR3) \cite{Planck:2018nkj}, and the fourth release (PR4), which has been processed by the NPIPE pipeline \cite{Planck:2020olo}. Both pipelines use the LFI and HFI raw and uncalibrated data
to generate frequency maps in the HEALPix \cite{Gorski:2004by} format, 9 in the case of temperature and 7 for Q and U Stokes parameters as the two highest frequency channels were not sensitive to polarization. These frequency maps are propagated through the component separation methods \cite{Planck:2018yye}, Sevem \cite{2012MNRAS.420.2162F}, SMICA \cite{2008arXiv0803.1814C}, NILC \cite{10.1111/j.1365-2966.2011.19770.x} and Commander \cite{2008ApJ...676...10E}, which generate each a clean CMB map. For PR3 we have used the data from the four separation algorithms. For PR4, clean CMB maps have been provided only by Sevem and Commander. We use both maps for the analysis in intensity and only Sevem products for polarization.\footnote{Although Commander polarization products are also available for PR4, 
we have not included them in this work, since the associated simulations were not well suited for our analysis (see Section \ref{PR4_E_Results}).} All the data and simulations are available at NERSC\footnote{National Energy Research Scientific Computing Center (NERSC), \url{https://www.nersc.gov/}, is a primary scientific computing facility operated by Lawrence Berkeley National Laboratory, located in California. It provides high-performance computing and storage facilities where Planck latest data and simulations can be found.} at full resolution\footnote{Data is also available at Planck Legacy Archive (PLA), \url{https:/pla.esac.esa.int/}.}  (N$_{\rm side} = 2048$), and have been downgraded (to N$_{\rm side}$=64) in harmonic space, according to, 
\begin{equation}\label{smoothing}
    a_{\ell m}^{out} = \frac{b_{\ell}^{out}p_{\ell}^{out}}{b_{\ell}^{in}p_{\ell}^{in}}a_{\ell m}^{in}
\end{equation}
where the superscripts $in$ and $out$ refer to the HEALPix resolution of the input and output maps, respectively, the $a_{\ell m}$'s are the harmonic coefficients of the considered maps, the $p_{\ell}$'s give the corresponding pixel window functions and the $b_{\ell}$'s are the Gaussian smoothing functions with a FWHM of 5$'$ for the original cleaned CMB maps and of 160$'$ for the downgraded maps. The output $a_{\ell m}$'s are then converted to a pixel map considering a maximum multipole $\ell_{max}$ = 3N$_{\rm side}-1$. For this procedure, we have used the Healpy package\footnote{\url{https://healpy.readthedocs.io/en/latest/}} \cite{Zonca2019}, which is basically a Python wrapper for HEALPix. Sevem maps for both
Planck releases
are shown in Figure \ref{fig:Maps}. Differences in polarization between both releases are apparent, mainly due to the reduction of large-scale systematics in the PR4 data. Conversely, intensity maps are very consistent.

All the details of the PR4 processing and the differences 
with respect to PR3
are given in \cite{Planck:2020olo}.
Most notably, while previous Planck pipelines process LFI and HFI in an independent way, PR4 was designed to process them simultaneously which produces a significant improvement in the calibration of both instruments. In addition, PR4 reduces the noise by including the data from the repointing maneuvers, which is around 8\% of the total mission. These differences, together with other added features and modifications,
have the net effect of reducing 
especially 
the polarization systematics at large scales 
as well as the level of noise. However, the signal at the smallest multipoles is also reduced due to the resulting transfer function (see section \ref{subsection:3.4} and Figure \ref{TF}).

In order to test the null hypothesis, we have used as the reference point the FFP10 PR3 and NPIPE Monte Carlo (MC) simulations, provided by the Planck collaboration along with the data maps. This set of simulated maps includes realizations of the CMB signal, the instrumental noise and the systematics, 
which try to capture the characteristics of the 
full data processing
such as the scanning strategy, the detector responses, or the calibration errors introduced during data reduction. For both PR3 and PR4, 
a number of realizations of the CMB sky are generated according to a fiducial CMB power spectrum based on the cosmology described in Table 6 of \cite{Planck:2020olo}. These sky realizations include lensing, Rayleigh scattering, and Doppler boosting. Additionaly, they are convolved by the frequency-specific beam. 
In contrast, noise simulations have been obtained following a more complex approach (see section 4 of \cite{Planck:2019evm} and section 5 of \cite{Planck:2020olo}), processing them using the same algorithm as for the real data via an \emph{end-to-end} (E2E) pipeline, including instrumental noise and systematics. For PR3, after propagating E2E simulations through component-separation pipelines, the final product is a set of 999 CMB and 300 noise simulations for each of the four methods. On the other hand, for the PR4 pipeline, 600 CMB and noise realizations are available for Sevem,
while 100 simulations are provided for intensity in the case of Commander.

Regarding masks, we use the PR3 intensity and polarization component separation confidence masks downgraded to $N_{side} = 64$, which retains the 71.3\% and 72.4\% of the sky, respectively.
Following \cite{Planck:2019evm}, to generate lower resolution binary masks, we first smoothed them according to Eq. \ref{smoothing} and then set a threshold of 0.95. All the pixels with a value below the threshold are set to 0, while the remaining ones are set to 1. Masks at $N_{side} = 2048$ are available in the Planck Legacy Archive\footnote{\url{https://pla.esac.esa.int/}} (PLA). The intensity mask is used
without modification for the local-variance analysis in the temperature maps. On the other hand, the polarization mask is used as the starting point for the inpainting of the Q and U maps and to generate a set of extended customised masks for the analysis of the E-mode map, as explained in section \ref{subsection:3.2}. 

\section{Methodology} \label{Met}

In this section, we first present an overview of our analysis and then explain in detail the different steps of the methodology in the corresponding subsections.
In the current work we perform an exhaustive analysis of the hemispherical power asymmetry anomaly for intensity and polarization following the methodology of \cite{Planck:2019evm}, where the local-variance estimator is used to find dipolar-like features. The estimator is introduced in subsection \ref{subsection:3.3}, and essentially the output is the amplitude and direction of the local-variance dipole, which can be use as a tracer of the modulation.

To apply this analysis to the polarization data, we first need to construct an E-mode map. This is a non-local quantity, and ideally we need full-sky maps of the Q and U Stokes parameters to have an accurate reconstruction of the E-mode map. However, since CMB observations are strongly contaminated by residual foregrounds in certain regions of the sky, we can not consider in the analysis those pixels excluded by the polarization confidence mask. To solve this problem, a useful approach is to carry out inpainting in the contaminated regions, a process that assigns new values to the excluded pixels such that they are more consistent with the rest of the CMB map. In particular, we have implemented a method based on \emph{Gaussian Constrained Realizations} (GCR) which is used to inpaint the regions masked by the polarization confidence mask in the Q and U data maps.
In subsection \ref{subsection:3.1} the inpainting method is described in more detail. 
We note that this technique requires the knowledge of the covariance matrix of all the components present in the map. For the CMB signal, 
this can be obtained analytically given a cosmological model, whereas for the noise and systematics component, we compute it from the E2E noise simulations. However, to avoid overfitting (due to the lack of convergence of the estimated matrix because of the relatively small number of simulations), 
the simulations considered in the analysis must be independent from those used to compute the covariance matrix. Therefore, we are forced to split the simulations in two halves, the first set used to estimate the covariance matrix and the second one to obtain the distribution of the amplitudes of the local-variance dipole and the corresponding p-value.\footnote{Note that the p-value is defined in an frequentist way as the fraction of simulations with an amplitude of the estimator equal or larger than the one obtained from the data.} Different partitions of the simulations are also considered in order to test the robustness of the results. In particular, subsection \ref{subsection:3.2} describes how the p-value is estimated taking into account the different data splits.

Another important step is to check the validity of the E-mode map constructed from the inpainted Q and U data maps. In particular, using the E2E simulations, we generate an extended customised mask by imposing a threshold in the pixel error of the reconstruction of the E-mode map. This mask is then used for the analysis of the polarization data (see subsection \ref{subsection:3.2} for details on how this mask is constructed).

Finally, section \ref{subsection:3.4} presents some tests to validate our methodology.
We check with the E2E simulations that the estimated amplitude and direction of the local variance dipole are not biased as a consequence of the inpainting process. We also give a comparison between the inpainting scenario and the one where E-modes are derived from a simple masking of the Q and U Stokes parameters. Furthermore, assuming a dipolar modulation model (Eq.~\ref{DM}), we obtain some interesting results such as 
the sensitivity of the method taking into account the realistic Planck noise and systematics.

\subsection{Inpainting using Gaussian constrained realizations}\label{subsection:3.1}

One of the main issues when analyzing the CMB is how to deal with foreground emissions such as dust, free-free, or synchrotron. Although component separation algorithms are able to reduce these emissions significantly, some residuals are expected to remain near the Galactic plane and in the
locations of point sources, where the sky is strongly contaminated. 
The standard approach is to mask these regions, 
but, as mentioned before, this leads to other  difficulties.
This is especially the case when dealing with polarization data,
since masking Stokes parameters introduces an undesired mixing between E- and B-mode \cite{Tegmark:2001zv, Lewis:2003an}. This effect has an impact 
when reconstructing the E-mode map, but it can be minimized by using an optimal inpainting technique. Inpainting consists on replacing the contaminated pixels by values which are somewhat consistent with the cleaned data, keeping as much as possible the statistical properties of the underlying field.
Different inpainting approaches have been considered in CMB analyses. For instance, 
diffuse inpainting, which consists on replacing the value of the contaminated pixels by averages of those of the neighbouring pixels, works acceptably well to inpaint small regions, such as those 
corresponding to point sources \cite{Planck:2018yye}. A second more sophisticated approach is purified inpainting, which constructs a full-sky polarization map from a masked one where most of the pure E- and B-modes are projected out in order to minimize the leakage. It has been used in the Isotropy and Gaussianity analyses performed by the Planck Collaboration \citep{Planck:2019evm}, including the study of the hemispherical power asymmetry.
In the present work, we use an alternative approach 
based on GCR. This technique was introduced in \cite{Marcos-Caballero:2019jqj} for temperature, but we have extended it here for the Stokes parameters.

GCR is a method that works in the pixel domain. The idea is to fill the masked pixels by sampling from the conditional probability distribution, $p(\mathbf{\hat{d}}|\mathbf{d})$, where $\hat{d}$ is the vector of the inpainted field and d is the vector of the available pixels. Assuming that the field is Gaussian, which is a very good approximation for the CMB, all we need is the pixel covariance matrix.
For the CMB signal, this 
can be computed from the theoretical power spectra\footnote{We use the $\Lambda$CDM best fit power spectra available in the PLA.} following, for instance, the Appendix A\footnote{
Note the small typo in equation A7 of the appendix
where the proportionality constant is negative, not positive. Thus, in equation A8, the negative sign corresponds to the case where z component of the vector $\hat{r}_{ij}$ is positive, and vice versa.} of \cite{Tegmark:2001zv}, whereas for the noise plus systematics part, we need to rely on simulations. We will present a full description of the method in a future work, together with a user-friendly
code (that will be publicly available), and some applications (Gimeno-Amo et al., in preparation). 
Although, in principle, this method is able to fill the contaminated pixels with a realisation that is perfectly consistent with the remaining data according to the assumed statistical properties of the underlying field, it presents two main limitations. First of all, this approach is very demanding from the computational point of view, since it requires the computation and storage of 
the pixel covariance matrix, whose dimension is $2N_{\rm pix} \times 2N_{\rm pix}$ when considering the inpainting of the Q and U Stokes parameters ($3N_{\rm pix} \times 3N_{\rm pix}$ if also TE correlation is considered). Hence the resolution at which we can work is
limited by the required memory. 
Due to this constraint, we have carried out our analyses at a Healpix resolution of $N_{\rm side}=64$. The second limitation comes from our lack of knowledge of the part of the covariance matrix coming from noise and systematics in the Planck polarization data, 
which needs to be characterised using simulations. In the case of the Sevem method for PR4, we have a total of 600 E2E simulations (600 for the CMB and 600 for the noise and systematics). As previously mentioned, to avoid overfitting\footnote{Overfitting arise from the non convergence of the covariance matrix. If we use the same simulations to estimate the matrix and perform the analysis, the matrix will have exactly the correlations given by the sample of simulations, and this will produce an overfitting when inpainting is applied on them.} 
in the inpainting process, we need to split them into two sets of 300 independent simulations, such that the first set is used to characterize the noise contribution to the covariance matrix, while the second one is used to obtain the distribution of the amplitude of the local variance dipole that will be compared with the data. 
The limited number of E2E noise simulations is not enough to achieve a good convergence of the noise covariance matrix, and consequently, matrix elements, especially the off-diagonal terms, are 
 significantly noisy. Indeed, using semi-realistic noise simulations -- Gaussian simulations that include anisotropic and correlated noise according to the noise covariance matrix obtained from the Planck noise E2E simulations -- we have found that the number of simulationes required to achieve convergence is at least of a few thousand. Another approximation is to ignore the --in general-- non-Gaussian behaviour of systematics, that is not taken into account in the inpainting. However, in spite of these limitations, 
 we find that this proposed inpainting procedure is still very useful for our purpose.
Indeed, it is not necessary to reproduce the full statistical properties of the inpainted pixels in order to minimise significantly the E-to-B leakage due to incomplete sky.

\subsection{Performance of the inpainting, confidence mask, and estimation of the p-value}\label{subsection:3.2}

\begin{figure}[t!]
    \centering
    \includegraphics[scale = 0.35]{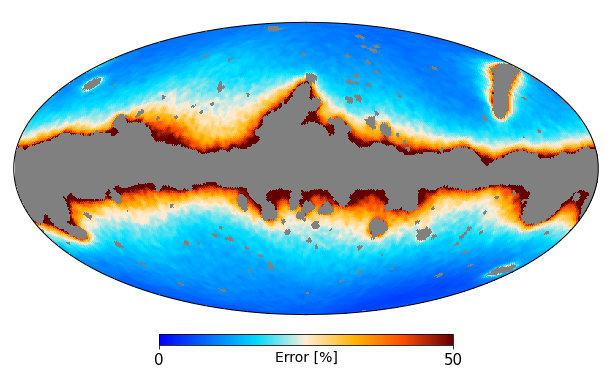}
    \includegraphics[scale = 0.35]{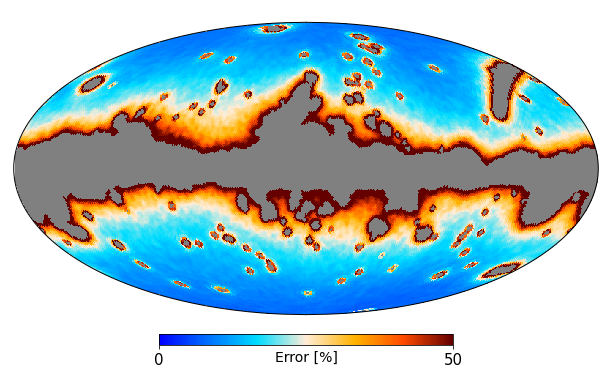}
    \caption{Comparison of relative error maps (in percentage) in the reconstruction of the E-mode for PR4. \textit{Left panel: }error map obtained with the inpainting approach. \textit{Right panel: }error map obtained for the simple Q and U masking approach (no inpainting). Grey area corresponds to the PR3 polarization confidence mask.}
    \label{errormap}
\end{figure}

The performance of the inpainting procedure can be determined by comparing the exact E-mode map obtained from full-sky Q and U maps of the E2E CMB plus noise simulations, with that obtained from the same Q and U maps that have been inpainted in the region given by the Planck polarization confidence mask.
In particular, we compute the map of E-mode residuals, $\Delta{E}$, as the difference between the exact and inpainted maps, for the pixels outside the mask for each of the inpainted simulations.
We also obtain the standard deviation of the residuals at each unmasked pixel ($\sigma_{\Delta E}$) as well as the one of the input E-mode map ($\sigma_E$). In the case of PR3, this procedure is repeated for the simulations propagated through the four component separation method, while for PR4 is done for Sevem. A relative error map $\delta{E}$ is then constructed as follows:
\begin{equation}
\delta{E}=\frac{\sqrt{\sum_m \sigma_{\Delta{E}}^2(m)}}{\sqrt{\sum_m \sigma_{E}^2(m)}}
\end{equation}
where $m$ runs over the different component separation methods used for PR3 or PR4. Note that for PR4, since only Sevem is considered, this is simply the ratio between the dispersion of the residuals and that of the input E-mode map at each pixel.
Figure \ref{errormap} shows a comparison between the map of the relative error in the E-mode map that we get by applying inpainting to the PR4 data (left panel) and by directly using masked Q and U maps (right). 
It becomes apparent that the inpainting works quite well, especially for point sources and isolated regions outside the Galactic plane. Moreover, in the ideal case in which the covariance matrix is perfectly known, we have found that the performance is even better. 

\begin{figure}[t!]
    \centering
    \includegraphics[scale = 0.35]{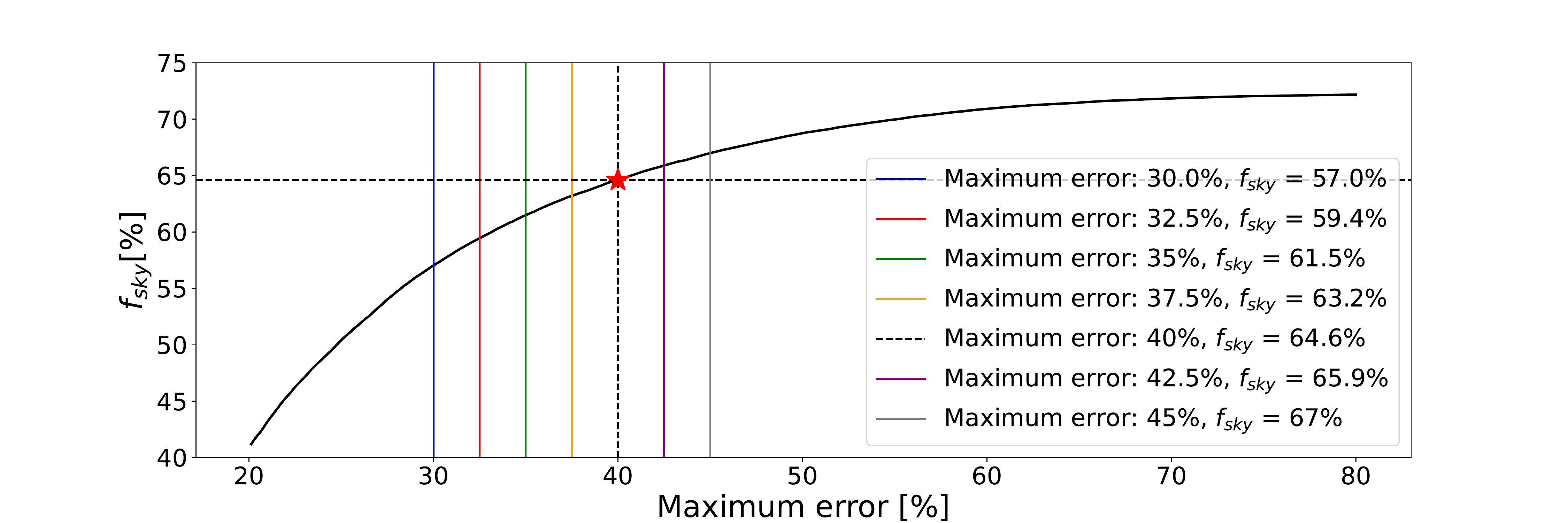}
    \caption{Sky fraction as a function of the allowed maximum error for the PR4 data set. 
    Vertical lines represent the thresholds for the different masks used in the analysis, covering from 45\% to 30\% error. The red star shows the maximum error and the fraction of the sky for our reference mask, which is used for the main results of this work and plotted in the right panel of Figure \ref{Mask}.}
    \label{FigMask}
\end{figure}

Despite having a more accurate reconstruction of the E-mode using inpainting, the residuals in regions close to the mask boundaries are still significant. One possibility to reduce these residuals is to extend the polarization confidence mask. In particular, we define a set of extended masks by selecting different thresholds in the error map, which are determined from a compromise between the maximum admissible error and the minimum 
extension of the mask. 
Figure \ref{FigMask} shows the variation of the maximum error with $f_{\rm sky}$\footnote{The sky fraction is actually determined from a combined mask obtained from the study of different splits. More specifically, the combined mask is obtained as the product of the individual extended masks generated for each of the considered splits (see below).} for PR4.
By removing the pixels more affected by the inpainting, an improvement in the maximum error is rapidly achieved, going down to values of $\delta{E}\sim$ 40\% with only a moderate reduction of the sky. However, further improvements require the removal of a significant fraction of the sky as a trade-off.
The complete set of masks will be used to check the robustness of the results against the considered sky fraction, but we select as reference mask the one with the threshold of $\delta{E}$ = 40\% because, for PR4, it leaves a fraction of sky similar to that of \cite{Planck:2019evm}.

This way of generating confidence masks is similar to the one followed in \cite{Planck:2019evm}, although with some differences. The first one is the resolution, since in that work the mask is generated at $N_{\rm side} = 1024$ and then downgraded to $N_{\rm side} = 64$, while we generate it directly at the final resolution (mainly due to our limitation to work with high resolution maps). 
The second difference comes from the way in which the threshold is chosen, since in their work the residuals of both the E and B-modes are considered.
We decided to focus only on the E-mode reconstruction for several reasons: B is consistent with zero in the Planck data, the E-mode signal is much larger than the B-mode, and finally, we are only applying the estimator to E maps.

Despite the increase in the error of the estimated dipolar modulation parameters due to the non-convergence of the noise covariance matrix, the amplitudes remain unbiased (see Section \ref{subsection:3.4}). However, splitting the simulations into two halves introduces additional uncertainty in the determination of the p-value. Then, the question is: how should we split the simulations? Although any random split is equally valid, each partition will yield a different p-value due to the slight variation in the amplitude distribution. There are also other effects that contribute to this dispersion but at a lower level. For example, the estimated noise covariance matrix is also different, which in turn affects slightly the way in which the inpainting is performed.

To get some additional insights on which is the best procedure to obtain the required p-value using the limited number of simulations, we have made some tests (see below) using semi-realistic noise simulations. These are constructed as Gaussian realisations characterised by the covariance matrix obtained from the full set of 600 E2E noise simulations\footnote{More specifically we calculate the Cholesky decomposition of the covariance matrix that is then used to generate the Gaussian realizations with the required statistical properties (see e.g. \cite{bar08}) for PR4.}. 
Therefore, these semi-realistic simulations of anisotropic noise contain the same correlations and statistical information up to second order as the realistic Planck simulations. The possible presence of non-Gaussianity is the only feature that we can not simulate. According to these simulations, a good estimation of the underlying p-value, the one that we would get if we sampled the distribution of amplitudes using the complete set of 600 simulations, can be achieved if we proceed as described below (this is detailed for PR4, but a similar procedure is applied to PR3):

\begin{figure}[t!]
\centering
  \includegraphics[scale = 0.35]{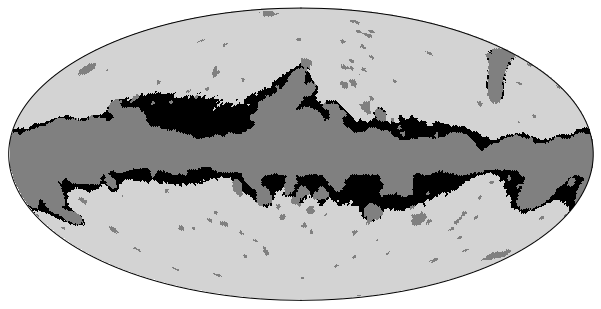}
  \includegraphics[scale = 0.35]{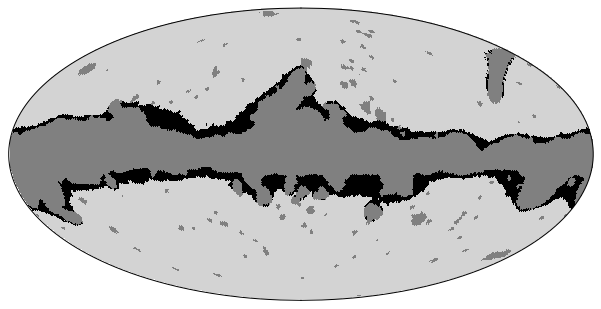}
  \caption{\textit{Left panel: }customized mask used for the PR3 data set analysis. It leaves 61.4\% of the sky unmasked. \textit{Right panel: }customized mask used for the PR4 Sevem data set analysis. It leaves 64.6\% of the sky unmasked. The grey region corresponds to the polarization confidence mask provided by Planck, while black region corresponds to the extended area where the relative error in the E-mode reconstruction  
  is larger than the 40\%.}
  \label{Mask}
\end{figure}

\begin{enumerate}
    \item We generate 30 random splits from the 600 E2E noise simulations. For each split we have one pair of independent sets of 300 simulations each, i.e., a total of 60 sets in total.
    \item For each split, we use the first set
    to estimate the noise covariance matrix and we apply inpainting on the second set.
    We repeat the same procedure but this time estimating the matrix with the second set and applying it to the first one.
    \item For each of the estimated covariance matrices (a total of 60), we construct an inpainted map from the real data.
    \item For each of the 60 inpainted sets, we generate an extended confidence mask by thresholding the residuals as explained previously for the reference case (i.e., maximum $\delta_E$ of 40 per cent). These are then combined into a single extended mask as the product of all of them. We repeat this procedure for different thresholds to have a set of extended confidence masks. Figure \ref{Mask} shows the reference extended masks for PR3 and PR4, used to obtain the main results of our analyses. 
    \item For each of the inpainted sets and inpainted data map (in each case inpainted using the same covariance matrix), and using the combined reference confidence mask, we apply the local-variance estimator (see section \ref{subsection:3.3}) to estimate the amplitude of the dipole. Finally, we define the p-value as the fraction of simulations with an amplitude equal or greater than the one observed in the data map.
    \item We get the mean of the 60 calculated p-values as our estimation of the underlying p-value. For the direction, we get the mean of 
    each of the vector components, and then, convert it to Galactic longitude and latitude coordinates.
\end{enumerate}

To establish the validity of the previous approach, the test we performed was the following one. 

\begin{enumerate}
    \item First, we generate 2000 CMB simulations modulated in Q and U according to equation \ref{DM}. We use an amplitude ($A_{DM}$) and direction consistent with the ones estimated from the data, $A_{DM} $ = 9\% and ($\ell$, b) = ($235^{\circ}, -17.5^{\circ}$).
    \item We add a semi-realistic noise simulation to each modulated simulation (we call them modulated set).
    \item We generate an independent set of 600 CMB simulations (unmodulated) and we add semi-realistic noise simulations, also independent from the previous 2000 realizations.
    \item Using as reference this set of 600 simulations, we estimate the p-value of each simulation in the modulated set. For this purpose, we apply the local-variance estimator (explained in section \ref{subsection:3.3}) to the ideal E-mode maps, reconstructed from the full sky Q and U maps (i.e., without applying inpainting). Just to clarify, we still apply an extended mask to the E-mode maps for the analysis. We label these p-values as the \emph{true} ones, since they are the best possible estimation obtained from the full reference set, i.e. using all the 600 simulations.
    \item Now let us focus on one of the simulations from the modulated data set. We treat it in the same way as we do for the data,
    and together with the 600 reference set we follow the steps described above, i.e., we generate 30 partitions of 300/300 simulations and we compute 60 p-values for the data (including the inpainting process as described previously). The same method is followed for the other 1999 simulations from the modulated set.
    \item We get the average value of the 60 p-values, finding that this estimator is a good representation of the previously estimated \emph{true} value. Taking the distribution of differences between \emph{true} and average values, we 
    compute the bias as the mean.
    As expected, the bias largely depends on the original p-value. In particular, if we restrict ourselves to simulations whose \emph{true} p-value is less than 8\% (in the real data the p-value varies between 2-3\%), the bias is below 1/300.\footnote{To test the robustness of this result, we have studied the bias using an independent set of 600 semi-realistic simulations, finding very consistent results.}
    \item Although the p-values obtained from different splits are strongly correlated as they are generated from the same set, we provide the range of p-values containing 68\% of the distribution as a rough estimation of the uncertainty.
\end{enumerate}

The same procedure has been followed for the PR3 data set, except that, in this case, we have only 300 E2E noise simulations to construct the 30 different partitions. Therefore, for each split, we have one pair of independent sets of 150 noise simulations each (used to construct the covariance matrix). In addition, we also have 999 CMB simulations, for which we consider 900 in our study. In this way, we construct sets of 450 simulations that are used for the analysis, obtained by combining one noise simulation with three different CMB simulations.

\subsection{Local-Variance estimator}\label{subsection:3.3}

As previously mentioned, the estimator we have used to characterize the asymmetry is the Local-Variance introduced by Y. Akrami \cite{Akrami:2014eta}. The main motivation is that a dipolar modulation of the anisotropies would manifest itself as a dipolar structure in a map generated by computing variances over sky patches.
The procedure to obtain the amplitude and direction of the dipolar modulation is the following:

\begin{figure}[t!]
    \centering
    \includegraphics[width=\textwidth]{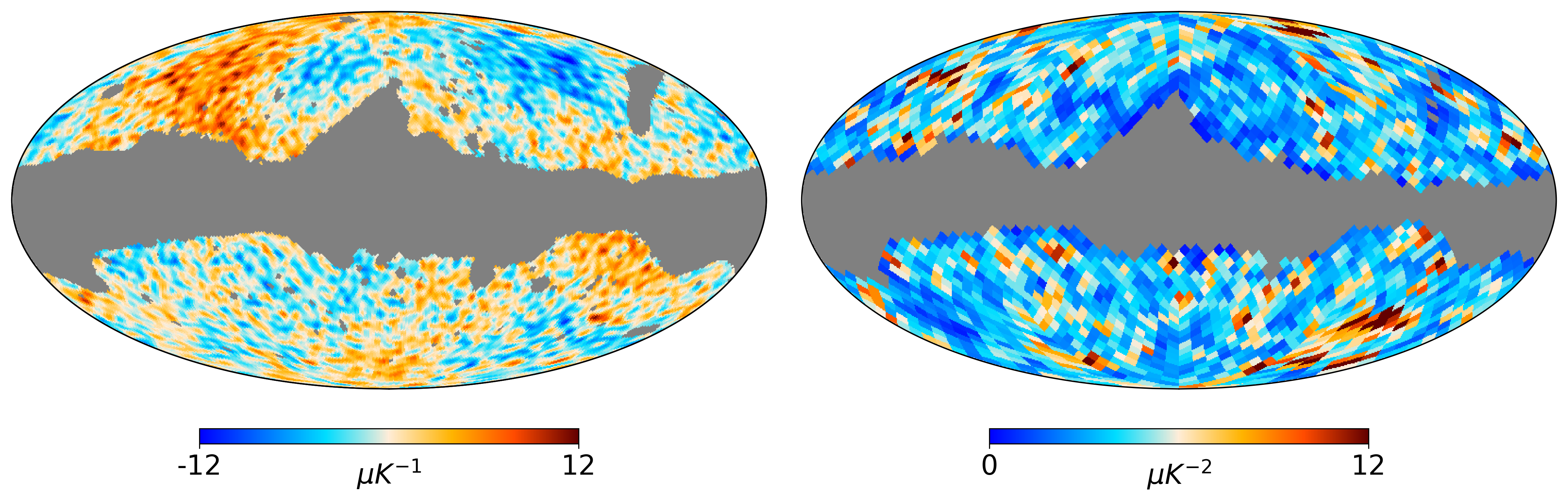}
    \caption{\textit{Left:} Sevem PR4 E2E E-mode simulation after applying the 
    transformation given by Eq.~\ref{eq:Bias}. The map is masked with an extended mask allowing a sky fraction of 63.2\%  \textit{Right:} Local-variance map generated following the previous steps. The grey region corresponds to those pixels discarded since more than 90 per cent of the pixels associated to the disc are masked. A disc of $4^{\circ}$ is used.}
    \label{fig:LVM}
\end{figure}

\begin{enumerate}
    \item This first step is exclusive to the analysis of polarization data due to the low signal-to-noise ratio. As mentioned in \cite{Planck:2019evm}, if we apply directly the method to the data, then the analysis returns local-variance dipoles whose directions are not uniformly distributed across the sky (see Figure 32 in \cite{Planck:2019evm}). Actually, their distribution is strongly correlated with the structure of the anisotropic noise. Therefore, we consider the following transformation directly applied on the input E-mode maps:
    \begin{equation}
        \label{eq:Bias}
        X^{'}_{i} = (X_{i}-M_{i})/\sigma_{i}^{2} 
    \end{equation}
    where $X_{i}$ is the value of the E mode map at pixel i computed from the full-sky inpainted Q and U maps. $M_{i}$ and $\sigma_{i}$ are the mean value and standard deviation at pixel $i$ obtained from inpainted simulations.

    \item We fix a specific HEALPix low resolution and we define a set of discs of a certain radius centered in the pixel centroids of the HEALPix map. Following previous works, we use $N_{side}$ = 16 and discs of 4 degrees for the present analysis. 
    
    \item We identify all the pixels of the map at the original resolution (in our case, $N_{side} = 64$ for polarization and $N_{side} = 64$ and 2048 for temperature)  that are inside each of the discs, compute their variance only taking into account the unmasked pixels, and associate these values to the pixel in the low resolution map. This is what we know as Local Variance Map (LVM). If more than 90\% of the pixels inside a disc are masked, then we mask the pixel in the LVM and remove it from the analysis. An example of one PR4 simulation is shown in Figure \ref{fig:LVM}.

    \item From the set of the LVM of the simulations, we estimate the mean $\bar{y}_i$ and the variance $\sigma_{d, i}^{2}$ on each disc, corresponding to one pixel at resolution $N_{\rm side}=16$. 

    \item Finally, we fit a dipole to each of the local-variance maps by applying a weighted $\chi^{2}$ 
    \begin{equation}
        \label{chi}
        \chi^{2}=\sum_{i}\frac{[(y_{i}-\bar{y}_{i})-d_{0}-\textbf{d}\cdot \hat{\textbf{r}}_{i}]^{2}}{\sigma_{d, i}^{2}}
    \end{equation}
    where $y_{i}$ is the LVM at pixel i; 
    $d_{0}$ captures the monopole component of the LVM; \textbf{d} = ($d_{x}$, $d_{y}$, $d_{z}$) is the dipole component; and $\hat{\textbf{r}}_{i}$ is the unit vector pointing to the i-th pixel. Sum is over all non-masked pixels in the LVM. $d_{0}$ and \textbf{d} are obtained by minimizing the $\chi^{2}$, which is analytic in this case.
\end{enumerate}

\subsection{Sensitivity and validation with simulations}\label{subsection:3.4}

\begin{figure}[t!]
    \centering
    \includegraphics[scale=0.4]{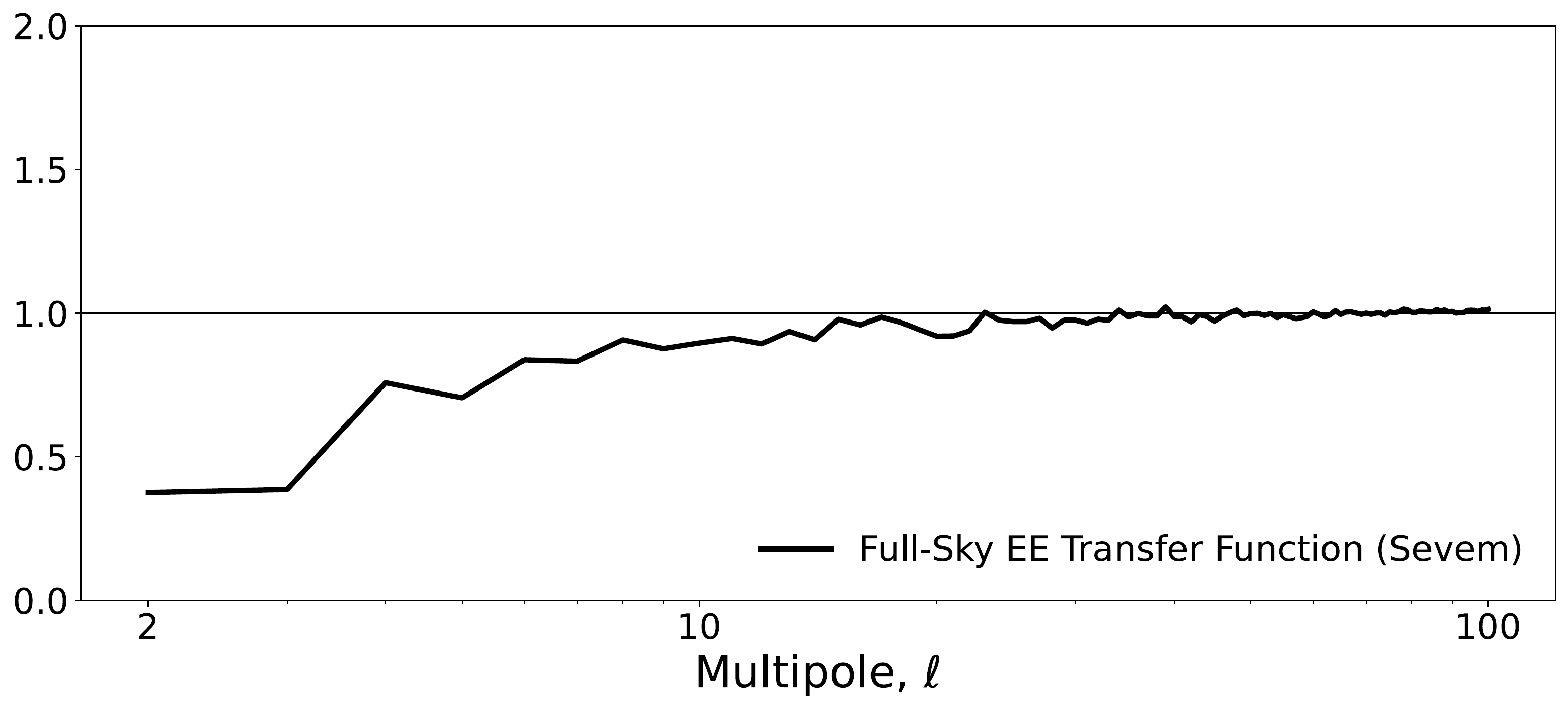}
    \caption{Sevem PR4 full-sky E-mode transfer function obtained from equation \ref{eq:TF}.}
    \label{TF}
\end{figure}

Before going into the results, we assess the sensitivity of the method by 
considering CMB simulations that contain a dipolar modulation, taking also into account the realistic E2E PR4 noise and systematics simulations. 
In particular, we remark that for PR4, a transfer function that affects only the lowest multipoles of the polarization is present in the data (see section 4.3 and Figure 20 from \cite{Planck:2020olo}). In order to take this effect into account for the E-mode map, we have computed the transfer function over the full-sky\footnote{Note that the transfer function is actually anisotropic, but using the full-sky transfer function is a sufficiently good approximation for our purpose.} for the Sevem data as: 
\begin{equation}
f^{EE}_{\ell} = \left( \frac{1}{N_{\rm sim}} \sum_{j=1}^{N_{\rm sim}} \frac{C_{\ell, j}^{\rm CMBx(CMB+N)}}{C_{\ell, j}^{\rm CMB}} \right)^2
\label{eq:TF}
\end{equation}
where $C_{\ell, j}^{\rm CMB}$ corresponds to the EE power spectrum of the $i$th PR4 Sevem CMB simulation and $C_{\ell, j}^{\rm CMBx(CMB+N)}$ to the cross-spectrum between that simulation and the same one after adding the associated noise and systematics Sevem simulation\footnote{The effect of the transfer function is actually included in the noise and systematics simulation, which presents therefore a correlation with its associated CMB simulation.}. Figure \ref{TF} shows the full-sky E-mode transfer function estimated for Sevem.

Starting with the E2E Sevem CMB simulations, we construct modulated maps using Eq. (\ref{DM}). Each simulation is modulated for 14 different amplitudes (going from 5\% to 18\%) and using a direction compatible with the one measured in the data (see section \ref{PR4_E_Results}), ($\ell$, b) = ($235^{\circ}, -17.5^{\circ}$). The modulation is applied to the Q and U maps, as a direct modulation in the E-mode map is not physically allowed \cite{Kothari:2018cmt}. The transfer function is then applied to the modulated CMB map\footnote{Starting at $\rm N_{side}$ = 2048, we downgrade the maps to $\rm N_{side}$ = 512 removing the Gaussian beam of 5$'$ and the pixel window function. Then, we modulate the Q and U maps, and we downgrade them to $\rm N_{side}$ = 64 applying to the $\rm a_{\ell \rm m}$ the square root of the Sevem E-mode transfer function and the Gaussian beam of 160$'$. We go through an intermediate resolution to avoid the numerical errors in the functions that convert from map to Fourier space.}. Finally, the E2E noise simulations are added to the simulated CMB signal\footnote{Note that the $i^{th}$ CMB simulation is added to an independent noise simulation (the $i^{th}$+1 one) to avoid including twice the transfer function.}.
All the tests presented here have been done using the customized mask with $f_{\rm sky}$ = 63.2\%. Although the specific results may be slightly different using other masks, the validation and main conclusions of this sub-section still hold.

The results of this sub-section have been obtained by combining a number of splits of the 600 modulated E2E simulations. In particular, as previously explained, we consider 30 different splits, a total of 30 pairs of 300 E2E noise simulations. Just to make it clearer, for each considered amplitude of the dipolar modulation we enumerate the most important steps of the procedure:

\begin{enumerate}
    \item For a given split, we have set A and set B, both of them with 300 CMB and noise simulations. We use one of them, for example set A, to compute the noise covariance matrix, and we apply inpainting on the set B. This inpainted set is what we call unmodulated analysis set.
    \item From the modulated E2E simulations we keep the ones whose associated noise simulations are not included in the set A. We apply inpainting on them, using the covariance matrix estimated from set A, and we treat them in the same way as the data (we call this the modulated data set).
    \item We compute the local-variance dipole amplitude and direction for the unmodulated set and for the modulated data set, and we estimate the p-value.
    \item Then, we can repeat this process but now using set B to estimate the covariance matrix. Additionally, we repeat the same procedure for the A/B sets of the other 29 splits. At the end, we get 30 p-values for each modulated E2E simulation\footnote{Note that for the case of the real data, since the data are independent of the simulations, we can obtain 2 p-values for each split (i.e. both data sets in the split can be used to construct the covariance matrix) and thus obtain a total of 60 p-values.}. This means that for each considered amplitude of the modulation, we have a total of 18000 local-variance amplitudes and p-values (30 splits x 600 modulated simulations), which are the values used to generate all the figures of this section. Although these values are not independent, they provide us with additional information with respect to use a single split and allow us to construct an approximated distribution of the estimated amplitudes.
\end{enumerate}
 
\begin{figure}[t!]
    \centering
        \includegraphics[scale = 0.37]{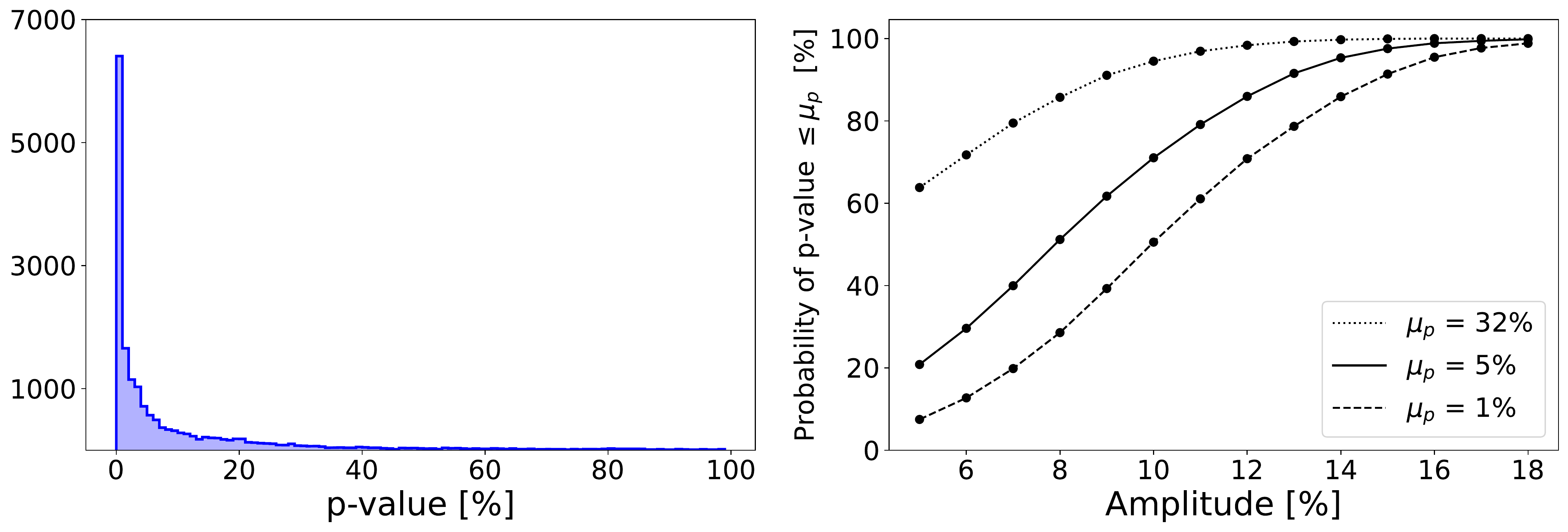}
    \caption{\textit{Left panel:} p-value distribution of the estimated amplitude of the dipolar modulation for the modulated simulations. We use an amplitude of 9\% and a fixed direction of ($\ell$, b) = ($235^{\circ}$, $-17.5^{\circ}$). \textit{Right panel:} probability curves as a function of the dipolar modulation amplitude with the realistic PR4 noise and systematics. 
    Assuming a dipolar modulation in the Stokes Q and U parameters, we could claim a detection with a p-value lower than 1\% (dashed line) in 95\% of the cases if the amplitude was larger than 16\%. Solid (dotted) lines correspond to the probabilities of getting a p-value equal to or lower than 5\% (32\%).}
    \label{Detectability}
\end{figure}

The left panel of Figure \ref{Detectability} shows the distribution of the p-value of the estimated amplitude of the dipolar modulation obtained as previously explained. The results correspond to simulations modulated with an amplitude of 9\%, which is close to the value found for the data (see section \ref{PR4_E_Results}). According to this figure, we would be able to find a p-value lower than 1\% in 39.3\% of the cases, while this percentage increases to 61.7\% for a p-value lower than 5\%. Additionally, the right panel shows similar information but for all the considered amplitudes. In particular, the percentage of times that we get
a p-value equal to or lower than 1\%, 5\% and 32\% for each amplitude is given. For Planck PR4 we find a p-value lower than 1\% in at least 95\% of the times for amplitudes larger than 16\%. The amplitude reduces to 14\% if we consider a p-value lower than 5\%. 
Using the set of modulated simulations we can also calibrate the method to give a relation between the local variance and the dipolar modulation amplitudes (see section \ref{PR4_E_Results}).

\begin{figure}[t!]
    \centering
    \includegraphics[scale = 0.4]{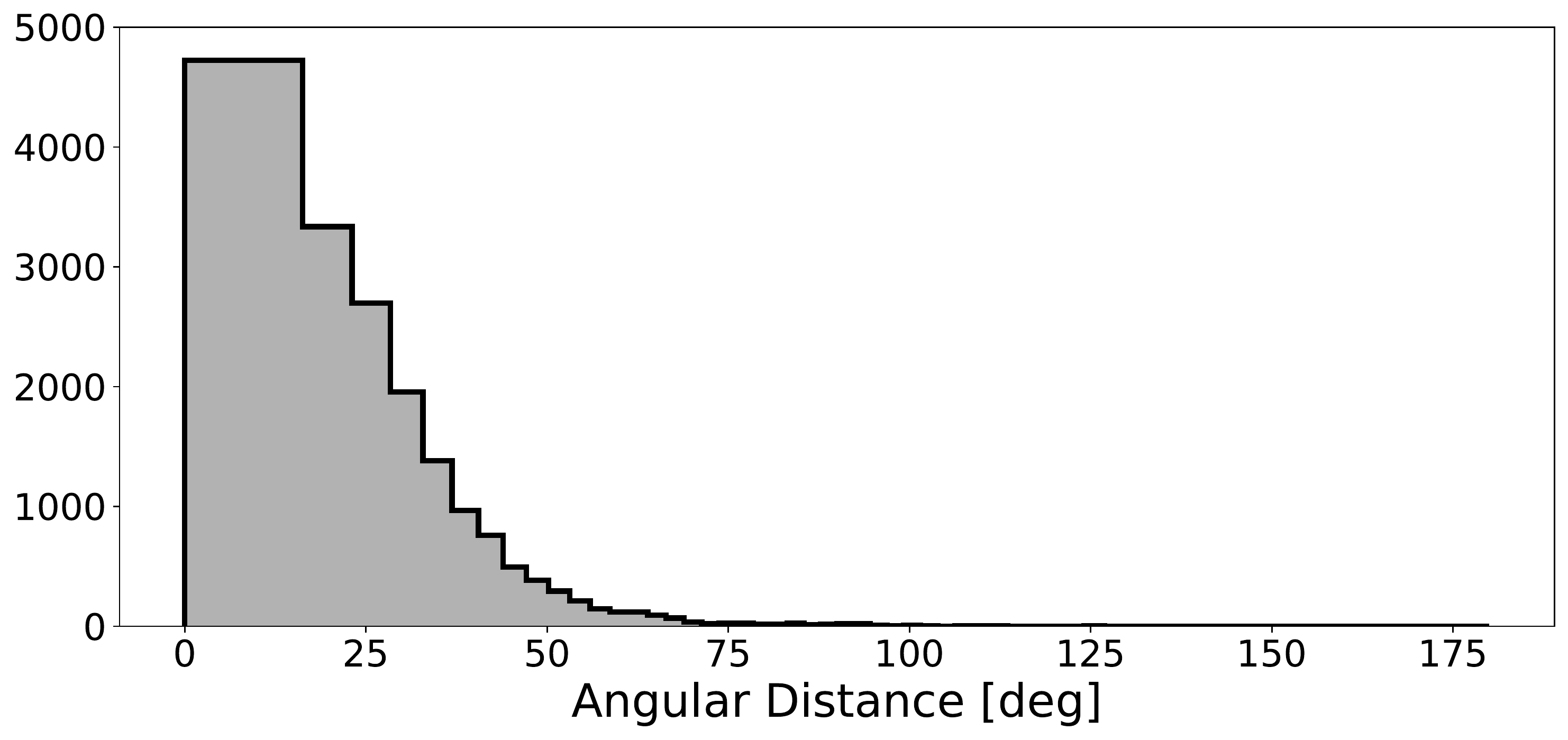}
    \caption{Distribution of angular distance between the input direction ($\ell$, b) = ($235^{\circ}$, $-17.5^{\circ}$) and the one obtained after applying the LVM estimator on the modulated ($A_{DM}$ = 9\%) PR4 simulations.  The ideal case, in which the Q and U full sky maps are known, is considered.}
    \label{AngularDistance_Sensitivity}
\end{figure}

We have also checked how well the method performs in estimating the direction of the dipolar modulation
by using the modulated simulations with an amplitude of 9\% (selected taking into account the results from Section \ref{PR4_E_Results}). In this case, we consider the ideal case of Q and U full-sky maps, to study the performance of the method itself, without including possible distortions introduced by the inpainting technique. We compare the longitude and latitude measured in the local variance maps for the 600 modulated simulations with the input values (fixed to ($\ell$, b) = ($235^{\circ}$, $-17.5^{\circ}$)), and compute the angular distance between them. Figure \ref{AngularDistance_Sensitivity} shows the distribution of the angular distance, where bins are not uniformly distributed, but instead they are selected so the area covered by each of them is constant. As seen, the dispersion in the estimated direction is quite large, even in this ideal case, although one should take into account that the considered amplitude is relatively small.

\begin{figure}[t!]
    \centering
    \includegraphics[scale = 0.42]{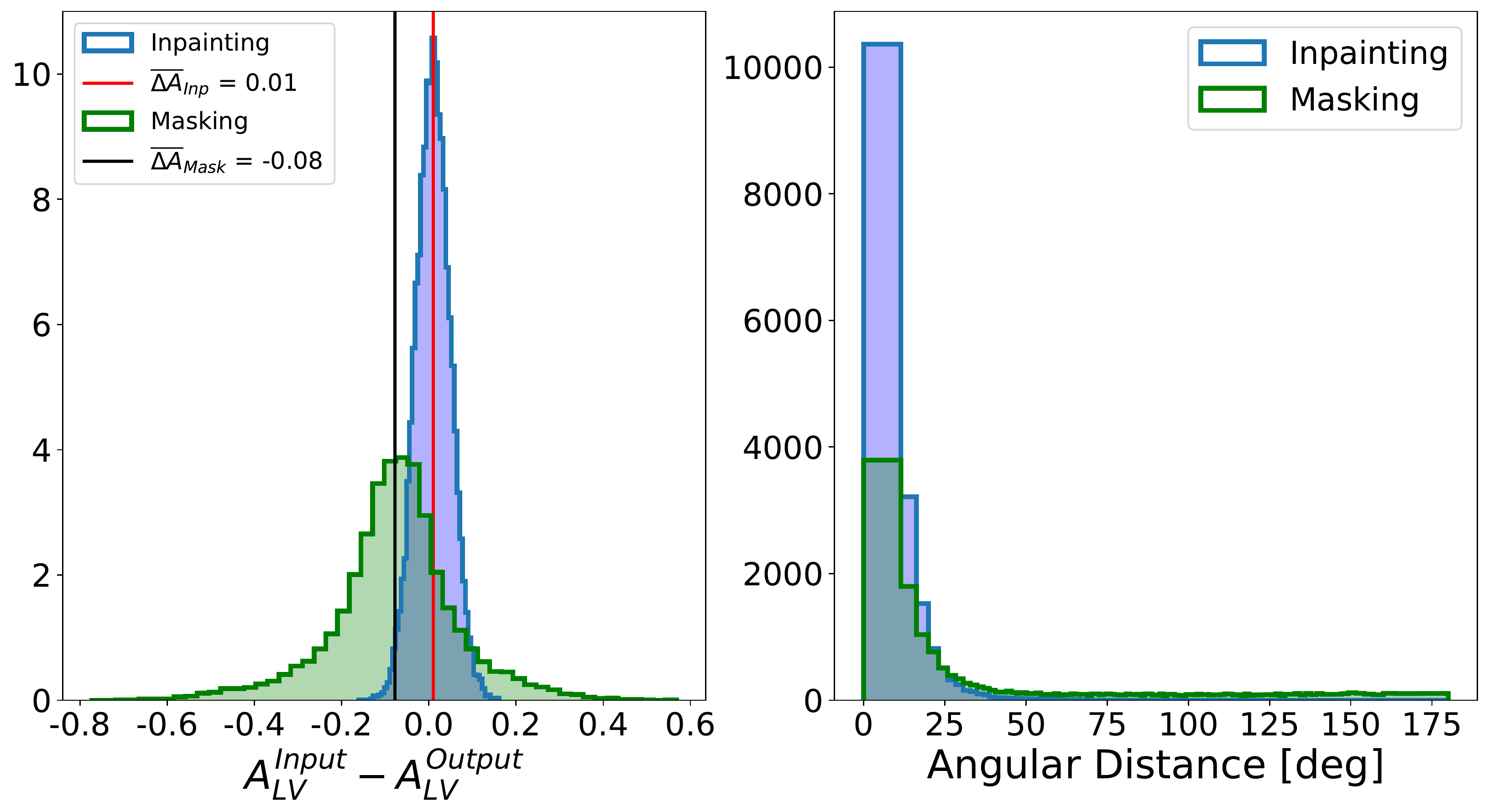}
    \caption{\textit{Left panel: }distribution of $\Delta{A_{LV}}$, defined as the difference between the amplitude measured on the inpainted (blue) or masked (green) simulations (denoted as output) and the one obtained when full-sky simulations are used (denoted as input). 
    Red and black lines correspond to the mean of the blue (inpainting) and green (masking) distributions, respectively. \textit{Right panel: }angular distances (in degrees) between the direction measured on the inpainted simulation and the input ones (blue). Input values are obtained from the corresponding full-sky (non-inpainted) simulations, so we remove the dispersion introduced by the estimator itself by making the difference. In green the same quantity is plotted for the case where a simple mask is applied on the Stokes parameters.}
    \label{Bias_A_Dir_And_Dist}
\end{figure}

We have carried out a final test to study the specific effect of the inpainting in both the estimated amplitude and direction of the dipolar modulation. For this, we have estimated those quantities from the 600 (unmodulated) E2E simulations in the ideal case in which Q and U full-sky are considered (that we call \emph{input} case). Then we have estimated the same quantities using inpainting. For comparison, a simple masking approach, where the masked area is simply replaced by zeroes in the Q and U maps, has also been considered. The left panel of Figure \ref{Bias_A_Dir_And_Dist} displays the distribution of the difference $\Delta{A}$ between the input amplitude and the one recovered from the inpainting or the simple masking approaches. It becomes apparent that the inpainting technique improves significantly with respect to simply masking the data. Indeed, on average, the latter tends to overestimate the amplitude, while no significant bias is found for the former. Furthermore, the dispersion of the distribution is significantly larger for masking (0.11) than the one we get when using our inpainting technique (0.04).

On the right panel, we also display the angular distances for the same scenarios. Again, we are not using uniform bins. From the figure it is clear that the inpainting provides a better reconstruction of the direction than the simple masking. We also note that although the dispersion introduced by the inpainting in the estimation of this quantity is significant, it is clearly below the intrinsic dispersion of the method (see Figure~\ref{AngularDistance_Sensitivity}). Furthermore, the median of the blue (inpainting) distribution is 10 degrees, while for the green one (masking) is almost 34 degrees.

\section{Results} \label{Res}

In this section, we show the results of the analysis of PR4 intensity and E-mode polarization data. For each case, we present the p-values and directions. We also re-examine the HPA in the PR3 data set using our alternative inpainting approach to check consistency with previous results.

\subsection{Intensity results for PR4}

Following previous temperature analysis, we present the results of full-resolution, $N_{side} = 2048$, PR4 data for a set of discs with radii between 4 and 40 degrees. The p-values are defined as the fraction of simulations with local-variance dipole amplitude larger than the one observed in the data. Unlike PR3, where cleaned maps were provided for four methods, for PR4, CMB maps are only available for Sevem and Commander. As expected, both of them are in good agreement with the Planck 2018 analysis. For Sevem, none of the 600 simulations have an amplitude as large as the one observed in the data for discs of 4, 6, and 8 degrees. This means that, under the $\Lambda$CDM model assumption, the probability of having such asymmetry in the temperature sky in that range of angular scales is below 0.17\%. For radii larger than $8^{\circ}$ the p-value increases systematically. We have checked with simulations that have a dipolar modulation pattern, with A = 7\% and p = ($209^{\circ}, -15^{\circ}$), that this increase of the p-value with the radius of the disc is expected. We obtain the same behavior for Commander, where none of the 100 simulations shows an amplitude as large as the one measured in the data. {Note that due to the lower number of simulations available for Commander, our sensitivity for the p-value is 1\% for this method.

Taking into account that the asymmetry only appears at large angular scales, we repeat the analysis for low resolution maps, at $N_{side} = 64$. Figure \ref{fig:T-results} shows the p-value as a function of the disc radius for both component separation methods and both resolutions. For low resolution maps the p-value is computed using discs with radii between 4 and 90 degrees as in \cite{Planck:2019evm}. In general, the significance level is lower than in the full-resolution analysis. Nevertheless, for discs of $4^{\circ}$ and $6^{\circ}$, we continue to find that no simulation has an amplitude equal to or greater than that of the data.

\begin{figure}[t!]
  \centering
  \includegraphics[scale=0.4]{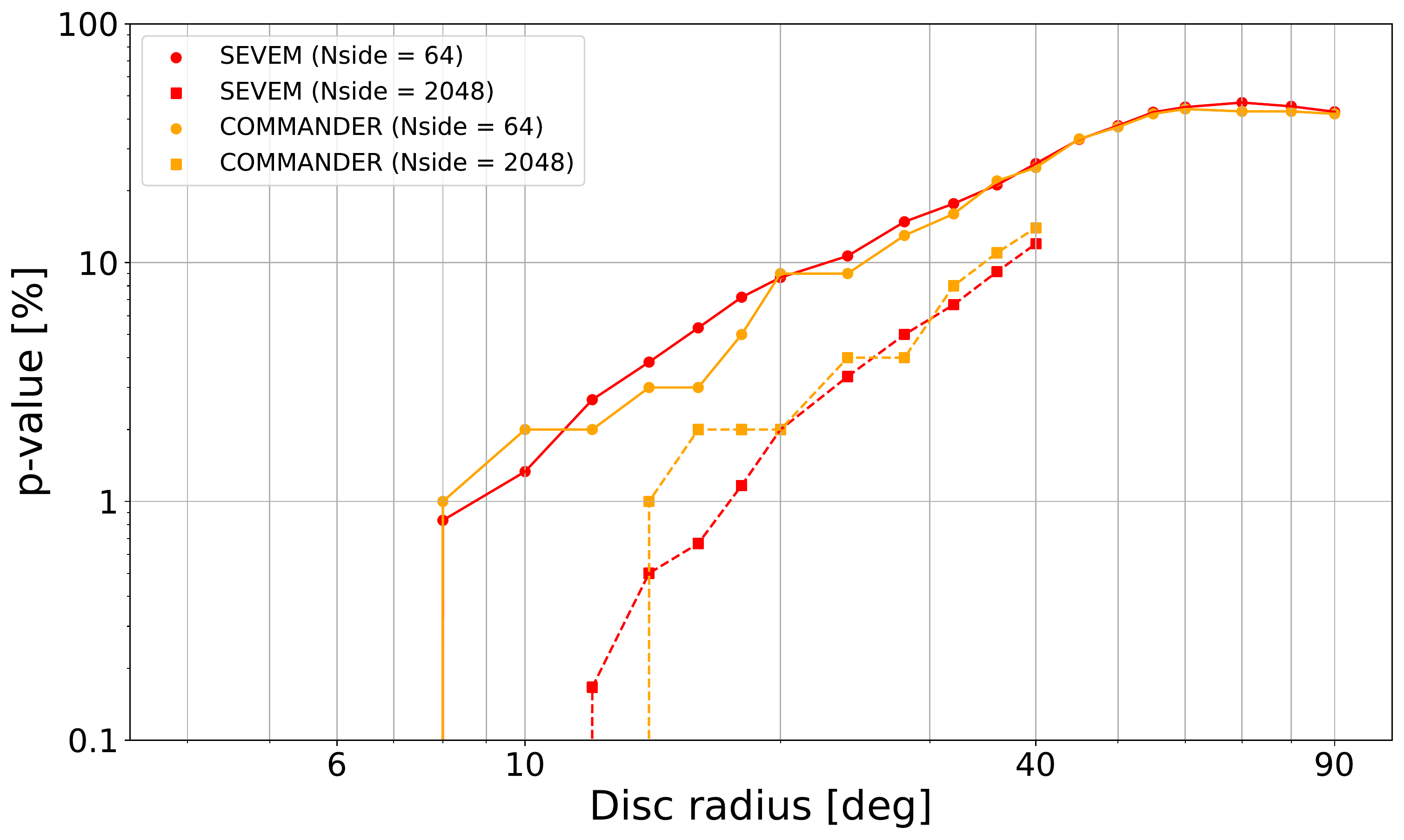}
  \caption{p-values for the asymmetry measured through the local-variance estimator for PR4 Sevem and Commander temperature maps and the two considered resolutions, $N_{side}$ = 2048 and $N_{side}$ = 64. The p-value is inferred by checking how many simulations have a local dipole amplitude larger than the one observed in the data. The sensitivity depends directly on the number of available simulations, corresponding to 0.17\% for Sevem and 1\% for Commander.}
  \label{fig:T-results}
\end{figure}

Regarding the direction of the dipole, in Table \ref{tab:dirT} we provide the full-resolution and low-resolution direction for Sevem and Commander obtained by considering $4^{\circ}$ discs. In Sevem the observed direction is in excellent agreement with the previously reported one (see Table 22 from \cite{Planck:2019evm}), while in Commander there is a small discrepancy of a few degrees, which could be due to the difference in the number of simulations between releases. In particular, we have checked that the $\ell$ and b angles have a very large dispersion (see Figure \ref{AngularDistance_Sensitivity}), especially when the amplitude of the modulation is small, making consistent the values for Commander between both releases.

\begin{table}[t!]
\centering
\begin{tabular}{ccccc}
\cline{2-5}
          & \multicolumn{4}{c}{($\ell$, b) {[}deg{]}}                                     \\ \cline{2-5} 
          & \multicolumn{2}{c}{PR4}               & \multicolumn{2}{c}{PR3}               \\ \hline
Data      & $N_{side}$ =  64 & $N_{side}$ =  2048 & $N_{side}$ =  64 & $N_{side}$ =  2048 \\ \hline
Sevem     & (208, -15)       & (205, -20)         & (209, -14)       & (205, -19)         \\
Commander & (213, -16)       & (207, -20)         & (209, -15)       & (205, -20)         \\ \hline
\end{tabular}
\caption{Local-variance dipole directions for PR4 Sevem and Commander temperature maps. Directions are measured in galactic coordinates for both resolutions and using $4^{\circ}$ discs.}
\label{tab:dirT}
\end{table}

\subsection{E-mode polarization results for PR3}

In order to compare the performance of our GCR inpainting technique versus the purified inpainting used in the Planck 2018 analysis \cite{Planck:2019evm}, we have first carried out the analysis of PR3 data using our method.
We have followed the procedure explained in Section \ref{subsection:3.2}. We recall that, in this case, each component separation algorithm has in total 300 noise E2E simulations each of them accompanied by 3 independent CMB skies. We have done 30 splits of two independent sets of 150 noise simulations and computed the covariance matrices as previously explained in section \ref{subsection:3.2}. Combining the 150 noise simulations with the 450 independent CMB skies, for the analysis we have available 60 sets of 450 simulations. 

Table \ref{PR3_Results} summarizes the results of this work. For each component-separation algorithm we provide the preferred direction of the local variance dipole and the p-values obtained using the reference mask. In addition, in Table \ref{Alignment_PR3} we give the angular distance between the temperature and E-mode dipoles, and the p-value associated with this alignment. 
As one would expect, we recover results quite similar to those obtained by the Planck collaboration, validating also our approach. We can distinguish two groups. While for Sevem and Commander the p-value of having such a large amplitude is below 1\%, for SMICA and NILC the probability is at the level of 3-4\%. Regarding the TE alignment, we are below 6\% for all the component separation methods. 
Again results are overall consistent between both analyses, although the result for NILC and SMICA becomes less anomalous in our case (p-values of 5.7 and 2.0 versus 1.9 and 0.9, respectively). These small deviations can be explained by the fact that our analysis presents some differences with respect to the previous one, such as the inpainting technique, a different mask or the use of splits.

\begin{table}[t!]
\centering
\begin{tabular}{ccccc}
\hline
Data      & p-value {[}\%{]} & p-value range {[}\%{]}       & ($\ell$, b) {[}deg{]} & p-value {[}\%{]} \\
& (this work) & & &  (from \cite{Planck:2019evm} )
\\ \hline
Sevem     & 0.22                         & {[}\textless{}1/450, 0.44{]} & (232, -9) $\pm 4$     & 0.4              \\
Commander & 0.70                         & {[}\textless{}1/450, 1.2{]}  & (222, -9) $\pm 4$     & 0.7              \\
SMICA     & 4.4                          & {[}1.8, 6.4{]}               & (225, -12) $\pm 4$    & 5.5              \\
NILC      & 3.4                          & {[}1.6, 5.0{]}               & (238, -16) $\pm 5$    & 5.8              \\ \hline
\end{tabular}
\caption{Local-variance dipole directions and p-values for the PR3 four component-separated E-mode polarization maps, analysed with the inpainting procedure, at $N_{side}$ = 64, together with the 1$\sigma$ interval obtained from the 60 sets as explained in Section \ref{subsection:3.2}. Measured directions are also showed. The error is estimated from the distribution of the angular distances between the mean direction and the directions of the 60 data sets, taking into account the contour, with azimuthal symmetry, that includes 68\% of the directions. All the values have been obtained using $4^{\circ}$ discs. Last column shows the p-value obtained in the previous work by the Planck collaboration \cite{Planck:2019evm}.}
\label{PR3_Results}
\end{table}

\begin{table}[t!]
\centering
\begin{tabular}{ccccc}
\hline
          & \multicolumn{2}{c}{TE alignment (this work)} & \multicolumn{2}{c}{TE alignment (from \cite{Planck:2019evm})} \\
Data      & $\cos{\alpha}$      & p-value {[}\%{]}     & $\cos{\alpha}$      & p-value {[}\%{]}     \\ \hline
Sevem     & 0.91                & 4.2                  & 0.86                & 6.9                  \\
Commander & 0.99                & 0.7                  & 0.99                & 0.9                  \\
SMICA     & 0.96                & 2.0                  & 0.99                & 0.9                  \\
NILC      & 0.88                & 5.7                  & 0.97                & 1.9                  \\ \hline
\end{tabular}
\caption{Angular distance of alignment between the preferred direction of the local variance dipole observed in intensity and the one for the E-mode polarization data. We use the mean directions presented in Table \ref{PR3_Results} to compute the angular distance, and the p-value is obtained using the E2E simulations combining 
all the results from different splits in a single distribution. The previous results by the Planck collaboration \cite{Planck:2019evm} are shown in the last column .}
\label{Alignment_PR3}
\end{table}

\begin{figure}[t!]
  \centering
  \includegraphics[scale=0.4]{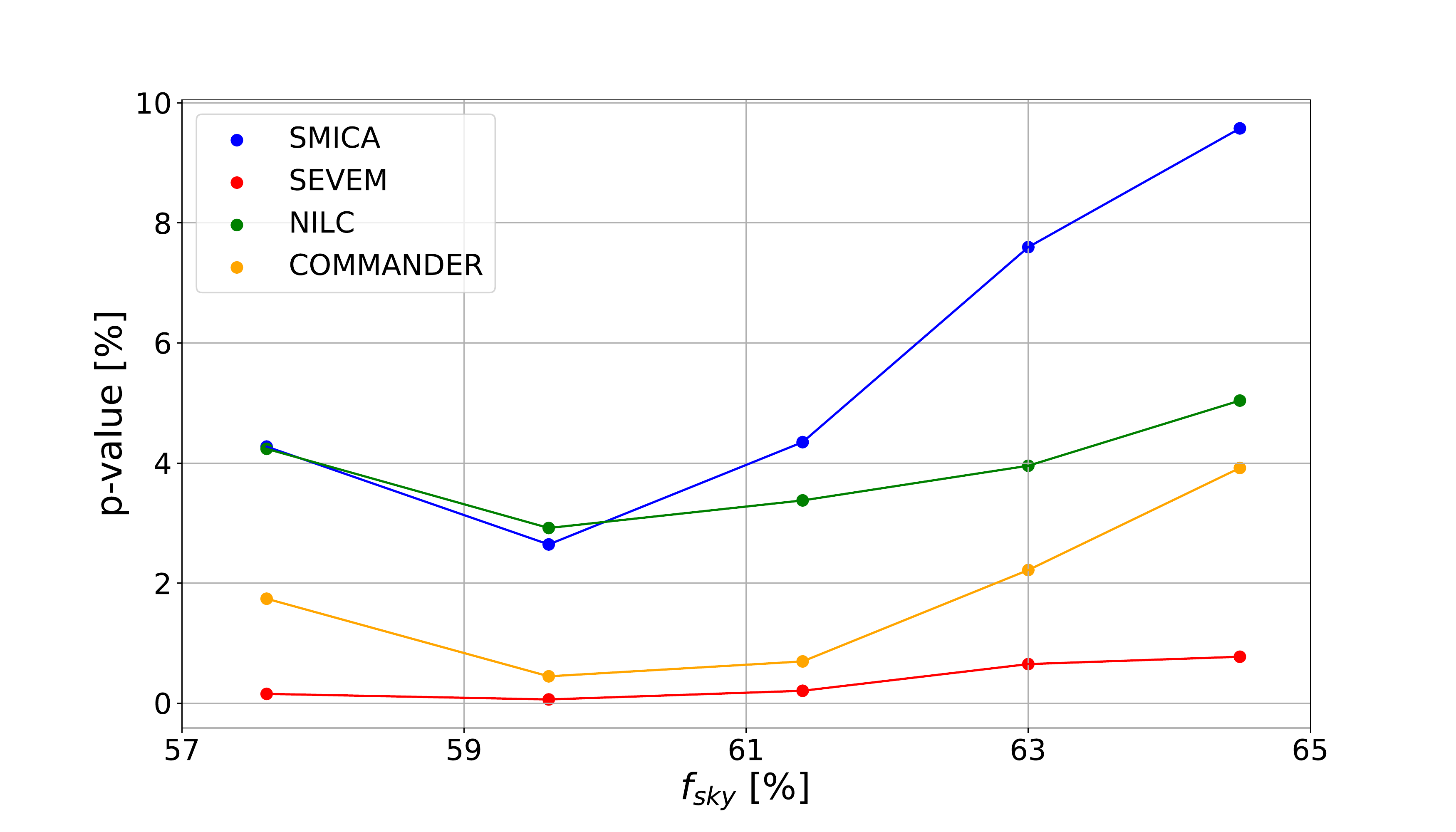}
  \caption{p-values for variance asymmetry as a function of $f_{\rm sky}$ for the four PR3 component-separated E-mode polarization maps analysed with the inpainting procedure.: SMICA (blue), Sevem (red), NILC (green) and Commander (orange). All values have been obtained at $N_{side} = 64$ and using $4^{\circ}$ discs. We recall that our reference mask for PR3 corresponds to a $f_{\rm sky}$=61.4\%.}
  \label{P_values_PR3_Against_Mask}
\end{figure}

\begin{figure}[t!]
    \centering
    \includegraphics[scale = 0.45]{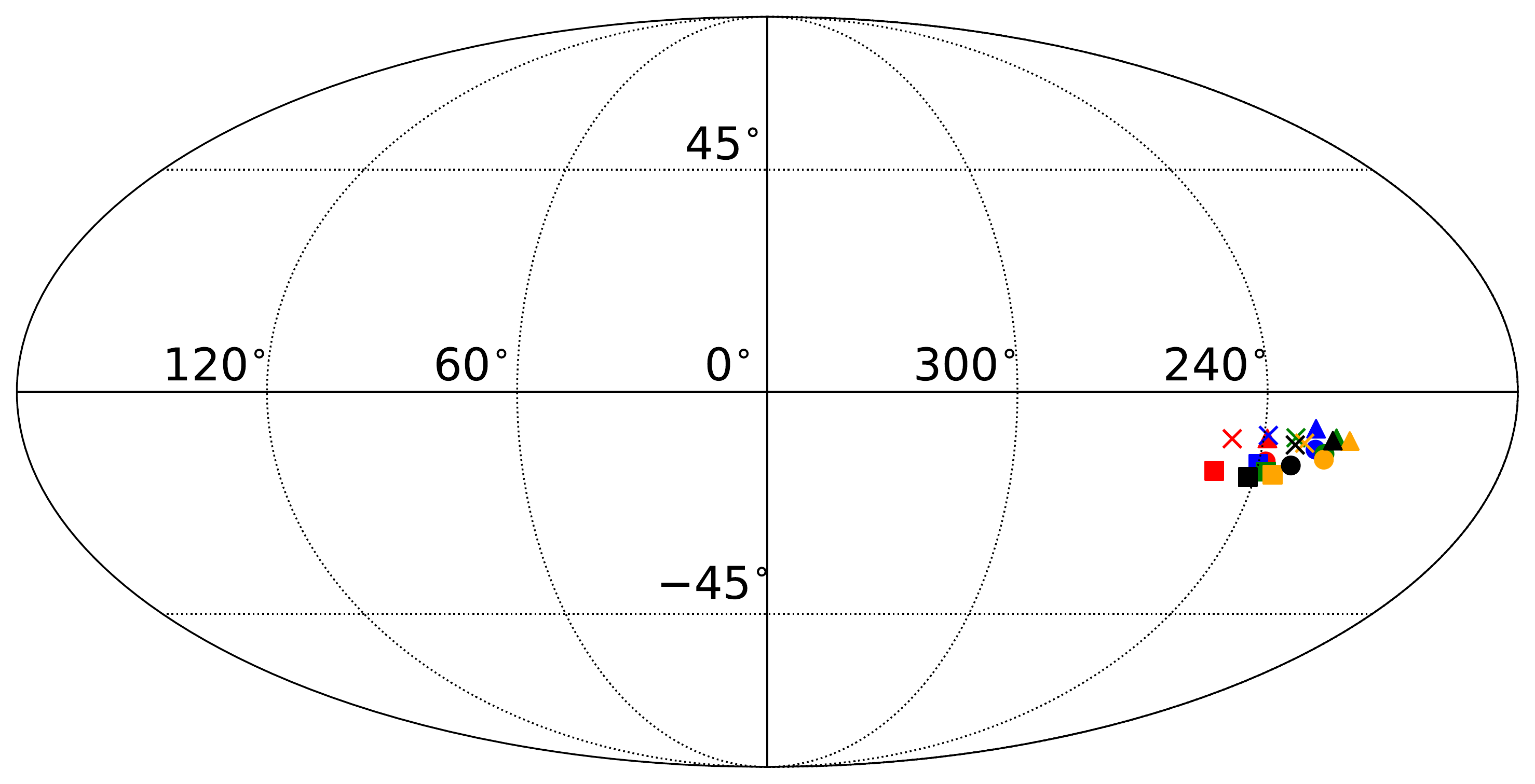}
    \caption{Local-variance dipole directions for the four PR3 component-separated E-mode polarization maps analysed with the inpainting procedure.: SMICA (circle), Sevem (cross), NILC (square) and Commander (triangle). Different colors corresponds to different $f_{sky}$: red (64.5\%), blue (63\%), green (61.4\%), orange (59.6\%), and black (57.6\%). All values have been obtained at $N_{side} = 64$ and using $4^{\circ}$ discs.}
    \label{Dir_PR3_Against_Mask}
\end{figure}

In order to test the robustness of the p-value against the mask, we have generated 4 additional masks using thresholds in the maximum reconstruction error (see Section \ref{subsection:3.2}) ranging from 45\% to 35\% and with the following allowed fractions of sky: 64.5\%, 63\%, 59.6\%, 57.6\%. Figure \ref{P_values_PR3_Against_Mask} presents the p-values obtained for the four component-separated E-mode polarization maps as a function of the $f_{\rm sky}$. The minimum p-value is obtained for the mask with $f_{\rm sky}$ = 59.6\%. Sevem and Commander are still below 1\% and the p-value is even lower than with the reference mask ($f_{sky} = 61.4\%$). However, SMICA and NILC present still values around or above a few per cent for all the considered masks. Additionally, Figure \ref{Dir_PR3_Against_Mask} displays the coordinates of the dipole direction as a function of the $f_{\rm sky}$. All the methods and masks show certain level of robustness in the sense that all the directions are concentrated in a small region near ($\ell$, b) = ($230^{\circ}$, $-13^{\circ}$). Furthermore, it seems that the latitude (b) is somehow more robust than the longitude ($\ell$).

\subsection{E-mode polarization results for PR4}\label{PR4_E_Results}

Finally, we present the results for the Sevem PR4 data set using the GCR inpainting \footnote{Note that we have not used the Commander PR4 data since CMB and noise simulations are not provided separately
(i.e. only CMB+noise simulations are provided), which makes not possible to compute independently the noise covariance matrix. 
Although, in principle, it would be possible to estimate the total covariance matrix, this introduces additional uncertainties in the matrix elements. Moreover, only 400 simulations are available for polarization, increasing even further the error in the analysis.}.
We have followed the same procedure as for PR3 with a total of 
30 splits of 300/300 simulations.

Table \ref{P_values_PR4} summarizes the results, providing the preferred direction of the local variance dipole and p-values obtained for the Sevem algorithm using the PR4 reference mask. The error in the direction is obtained as the 68 per cent (single-tailed) of the distribution of the angular distance between the mean direction and the direction of the 60 data sets. Therefore, it is only given to provide an estimation of the dispersion in the different considered splits. In practice the error associated to the estimated direction should be significantly larger, since this quantity is below the errors inferred from the methodology itself or the effect of the inpainting. 
We also give in the same table the angular distance between the temperature and E-mode dipole directions, and the p-value associated with this alignment. The analysis shows that the p-value of the amplitude for Sevem PR4 increases with respect to the one obtained for PR3 (see Table \ref{PR3_Results}), increasing from 0.22 to 2.8 per cent. The p-value of the TE alignment is very similar for both pipelines and remains below 5 per cent.

\begin{table}[t!]
\centering
\begin{tabular}{ccccc}
\hline
p-value {[}\%{]} & p-value range {[}\%{]} & ($\ell$, b) {[}deg{]} & 
\multicolumn{2}{c}{TE alignment} \\ 
                &                        &                       &      $\cos{\alpha}$ & p-value {[}\%{]} \\ \hline
2.8              & {[}1.8, 3.8{]}         & (234, -14) $\pm$ 5    & 0.91           & 4.5              \\ \hline
\end{tabular}
\caption{Local variance dipole direction and p-value for the Sevem PR4 E-mode polarization map, analysed with the inpainting procedure, at $N_{\rm side} = 64$, together with the $1\sigma$ interval obtained from the 60 data sets.
Measured directions are also showed. The error is estimated from the distribution of the angular distance between the mean direction and the directions of the 60 data sets,
taking into account the contour, with azimuthal symmetry, that includes 68\% of the directions. All the values have been obtained using $4^{\circ}$ discs.}
\label{P_values_PR4}
\end{table}

Another interesting result is the level of the modulation in the data. Using the modulated simulations described in Section \ref{subsection:3.4}, the relation between the amplitude of the modulation (given by equation \ref{DM}) and the one measured in the local-variance map can be estimated. Figure \ref{Calibration} shows this relation together with the amplitude observed in the data local-variance map for the reference mask. 
According to this, for modulations with amplitudes larger than 6\%, the relation is linear, although with a large dispersion due to the E-mode noise level. Assuming that the modulation model (Eq.\ref{DM}) is the correct one, the asymmetry observed in the data corresponds to a modulation at the level of 9\%. Moreover, the estimated amplitude $A_{\rm LV}$ is within the 68\% of the distribution for the dipolar modulation amplitude ($A_{\rm DM}$) for the range between 6\% and 13\%.
\begin{figure}[t!]
    \centering
    \includegraphics[scale = 0.4]{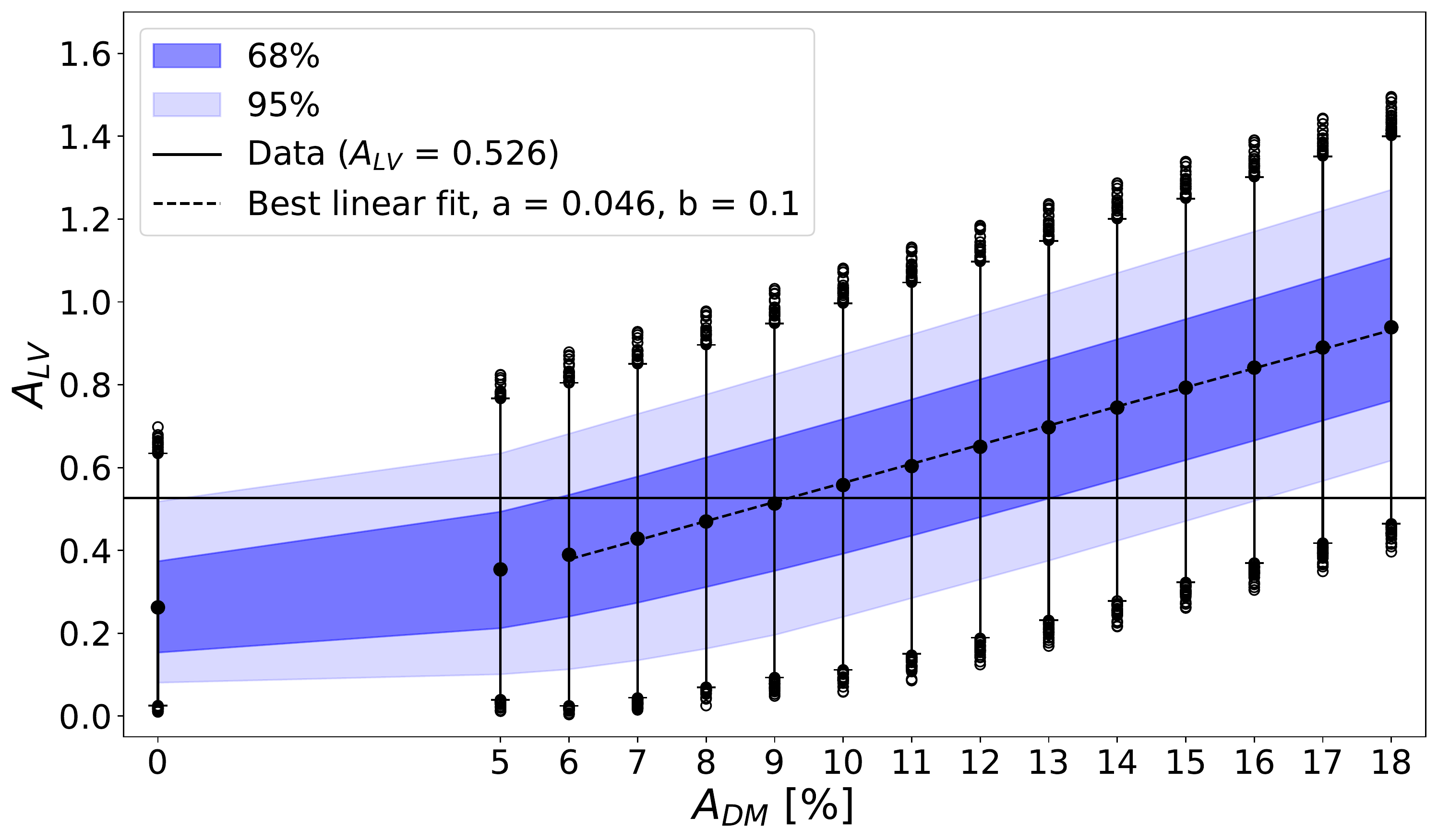}
    \caption{Relation between the dipolar modulation amplitude $A_{\rm DM}$ (input value in the modulated simulations) and the one measured in the local variance map $A_{\rm LV}$, assuming the realistic PR4 noise and systematics.
    For each amplitude, all the 60 inpainted data sets
    are used to generate a single distribution. For amplitudes larger than 6\%, there is a linear relation between both amplitudes with a slope of 0.046 (dashed black line). The horizontal black line corresponds to the mean amplitude (over the 60 data sets) measured in the data local-variance map using the PR4 reference mask. The measured $A_{\rm LV}$ value corresponds to
    a modulation at the level of 9\%, and it is within the 68\% of the distribution for the $A_{\rm DM}$ interval between 6\% and 13\%. Black dots in the tails of distributions represent the values outside the 99.7\%.}
    \label{Calibration}
\end{figure}

\begin{figure}[t!]
  \centering
  \includegraphics[scale=0.4]{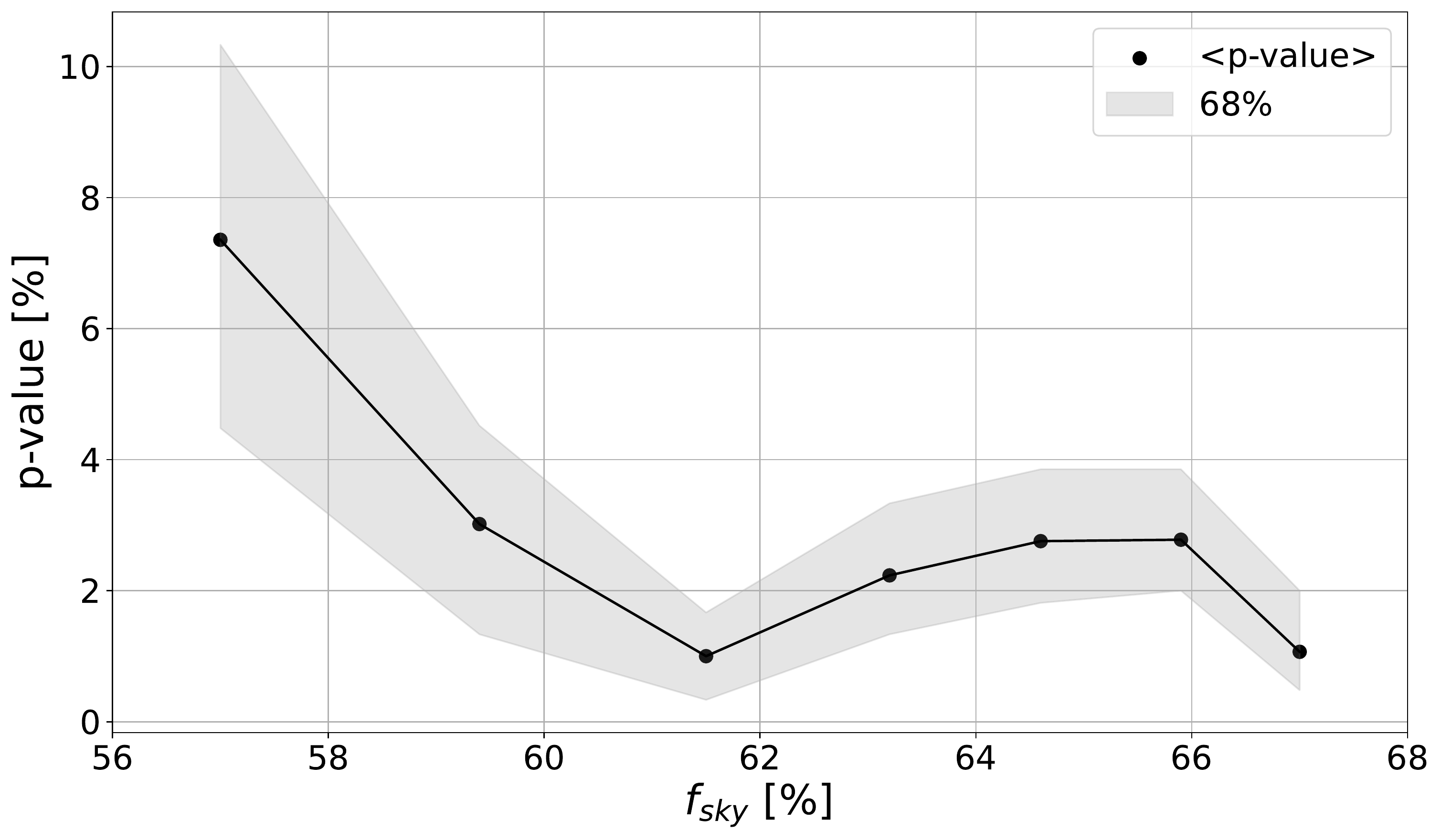}
  \caption{p-values for variance asymmetry as a function of $f_{\rm sky}$ for the PR4 Sevem E-mode polarization map analysed with the inpainting procedure. The grey region corresponds to the 68\% interval obtained from the 60 data sets. All values have been obtained at $N_{\rm side} = 64$ and using $4^{\circ}$ discs.}
  \label{P_values_PR4_Against_Mask}
\end{figure}

As before, we test the robustness of the p-value against the different masks. We have generated 6 additional masks using thresholds in the range from 45\% to 30\% with the following allowed fractions of sky: 67\%, 65.9\%, 63.2\%, 61.5\%, 59.4\%, and 57\% (we recall that the reference mask corresponds to f$_{\rm sky}$ = 64.6\%}).
Imposing more strict conditions in the error of the recovered E-mode map (i.e. lower thresholds) while keeping a reasonable $f_{\rm sky}$ is possible in this case because a more accurate E-mode reconstruction can be obtained in PR4. This is due to both, a lower level of systematics and twice the number of simulations that have been used for estimating the noise covariance matrix. Figure \ref{P_values_PR4_Against_Mask} presents the obtained p-values as a function of $f_{\rm sky}$. The minimum p-value (1\%) is obtained for the mask with $f_{\rm sky}$ = 61.5\%. For the 5 smallest masks, allowing a sky between 67\% and 61.5\%, the p-value is quite stable with a value below 2.8\%, and then it starts to rise with $f_{\rm sky}$. The largest mask returns a p-value of 7.4\%. 
This behaviour could be due to the loss of information as the mask increases, assuming the asymmetry had a cosmological origin.
However, it could also be due to the fact that a larger mask reduces foreground contamination, what we would expect if the asymmetry were due to the presence of residuals.
Additionally, Figure \ref{Dir_PR4_Against_Mask} and \ref{P_value_Alignment} display the directions of the local-variance dipoles and the p-value of the TE alignment, respectively, as a function of the $f_{\rm sky}$. 
In particular, the minimum p-value (4.6\%) is obtained for the reference mask. Finally, Figure \ref{Summary_Dir} summarizes in a Mollweide projection all the directions measured in this work.

\begin{figure}[t!]
    \centering
    \includegraphics[scale = 0.45]{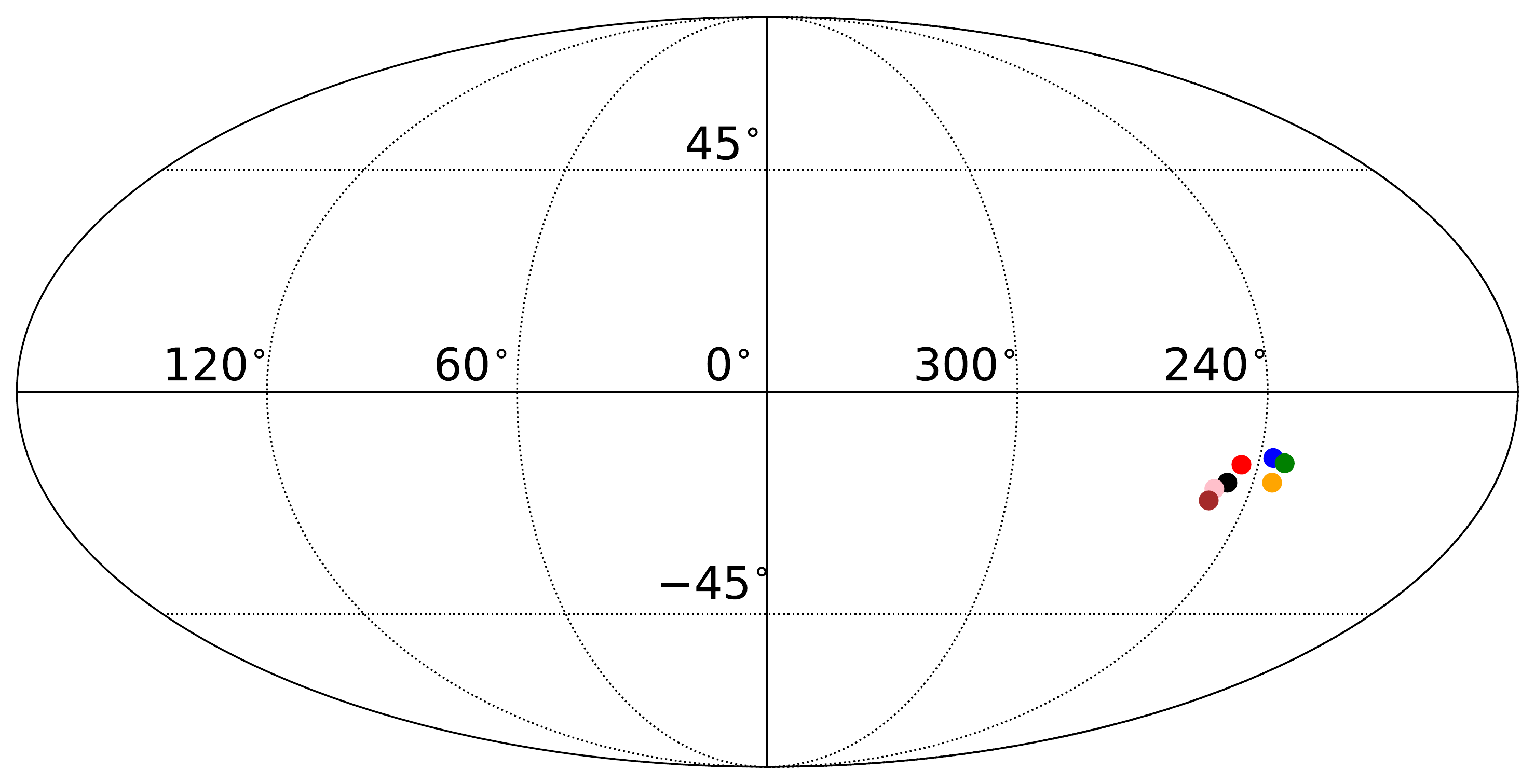}
    \caption{Local-variance dipole directions for the Sevem PR4 
    component-separated E-mode polarization maps analysed with the inpainting procedure. 
    Different colors corresponds to different f$_{\rm sky}$: red (67\%), blue (65.9\%), green (64.6\%), orange (63.2\%), black (61.5\%), pink (59.4\%), and brown (57\%). All values have been obtained at Nside = 64 and using 4$^\circ$ discs}
    \label{Dir_PR4_Against_Mask}
\end{figure}

\begin{figure}[t!]
    \centering
    \includegraphics[scale = 0.4]{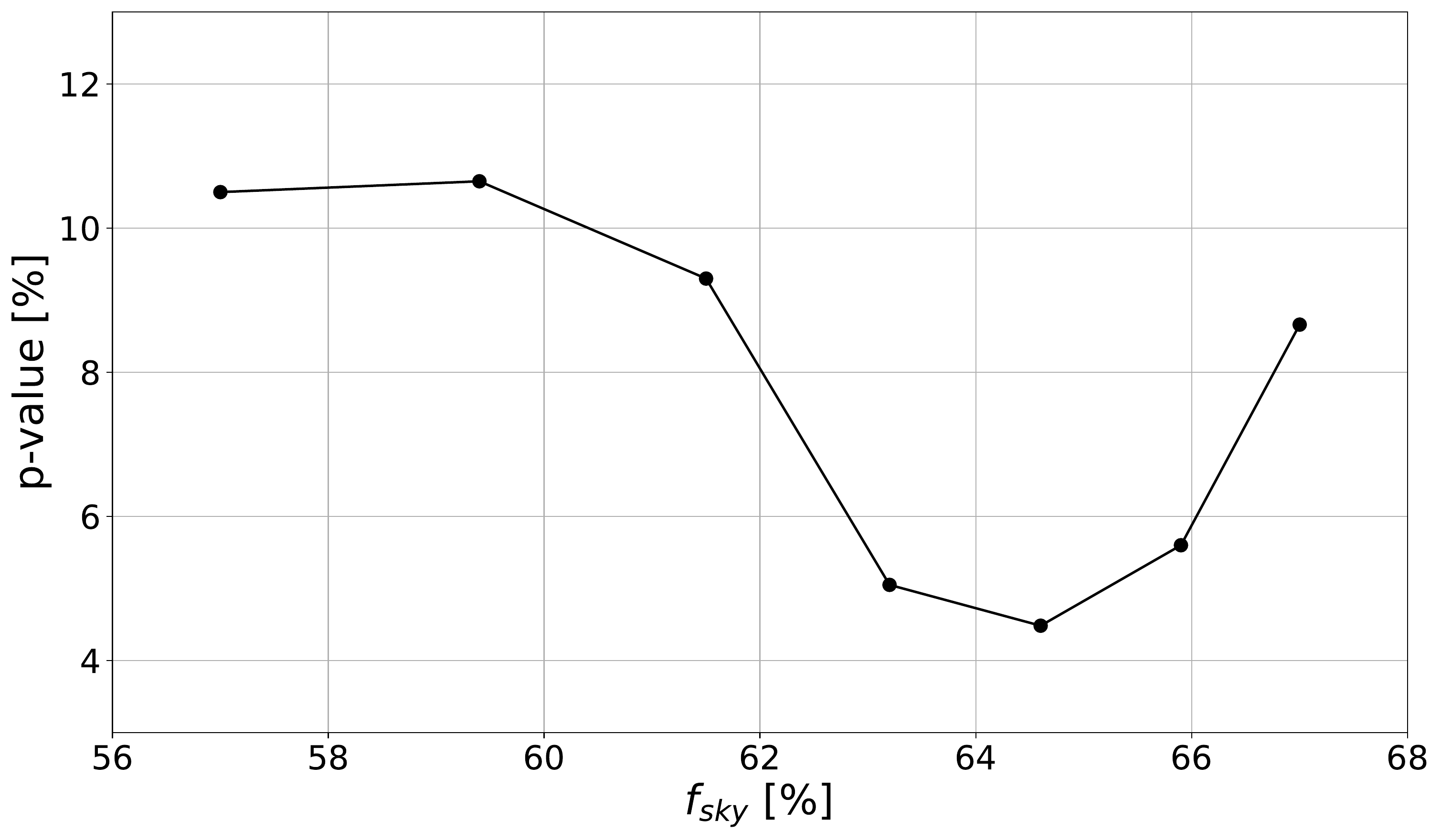}
    \caption{p-value of the alignment between the dipole axis in PR4 temperature and E-mode as a function of $f_{\rm sky}$.}
    \label{P_value_Alignment}
\end{figure}

We have performed a final test 
where we studied if the difference observed in the p-value from the PR3 and PR4 processing is consistent with that expected due to differences in the number of simulations, in the mask, inpainting, TF and noise properties. For this purpose, we construct two data sets that have the same input CMB sky but different noise properties. In particular, the first set is formed by the modulated 600 PR4 CMB simulations, plus the corresponding CMB noise (including thus also the TF). The second set consists of the same modulated PR4 simulations but adding the PR3 noise (repeating two times each simulation, because only 300 are available). Then, we run the full procedure, using the corresponding reference mask for each data set, and we compare the output p-values. Note that for the PR3 case, we have used as reference the PR3 CMB plus noise simulations, i.e. 450 simulations (450 CMB simulations and the 150 noise simulations, repeated three times, that are not considered in the matrix estimation).
The distribution we get from the 600 simulations for the difference between the p-value for PR4 minus the one for PR3 is peaked at -8.9\%, which means that the dipolar asymmetry can be better constrained using the PR4 processing even in the presence of the TF. The difference observed in the data (from 0.22\% to 2.8\%) is well within the 68\% CL (which corresponds to the range -34.9 \% to 8.0 \%), and thus 
both values are fully consistent taking into account the differences in the data and in the analysis. Finally, we also see large tails in the distribution, which is an indication of the low signal-to-noise ratio as the amplitudes and p-values are significantly affected by noise and systematics.

\begin{figure}[t!]
    \centering
    \includegraphics[scale = 0.45]{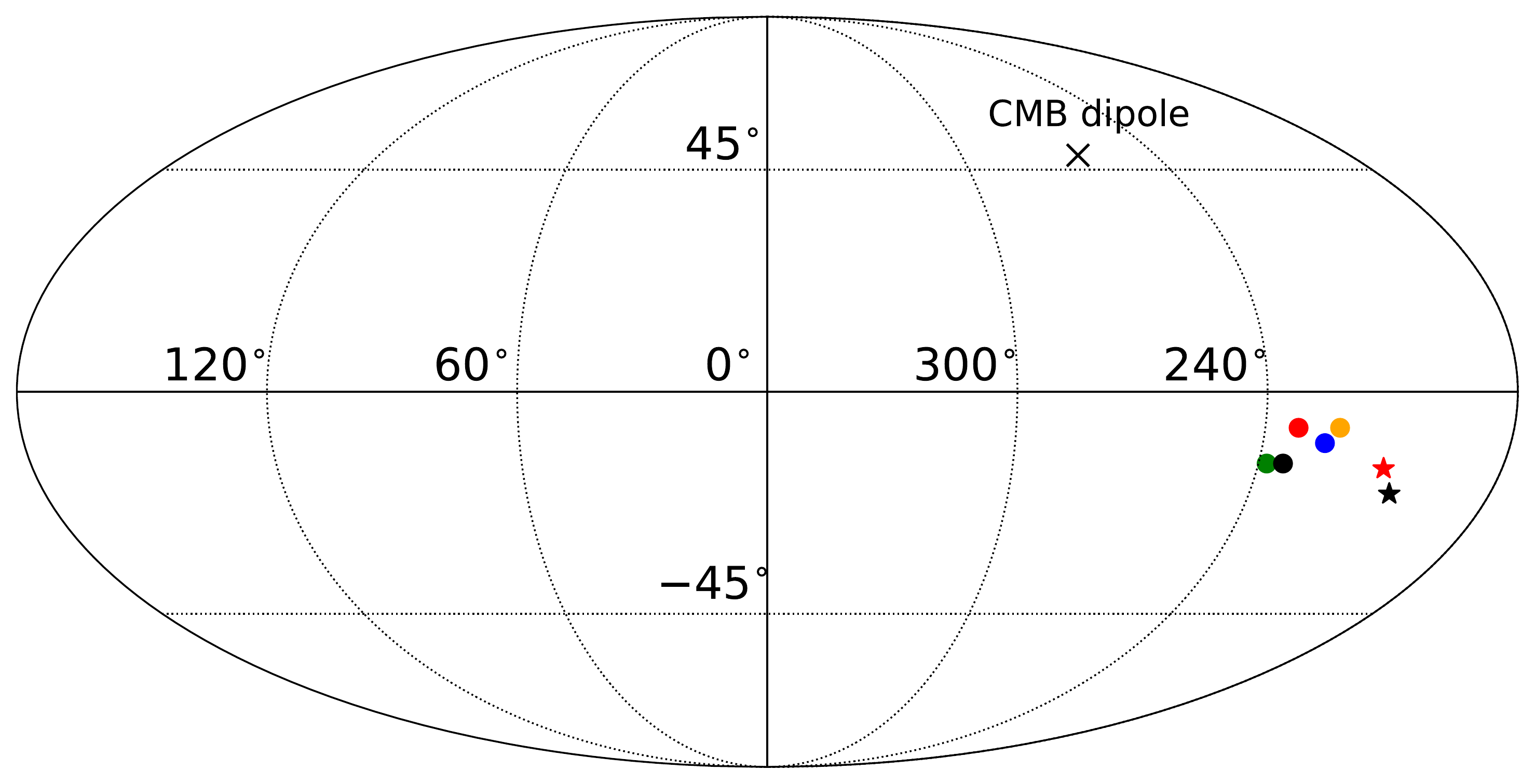}
    \caption{Local variance dipole directions for the PR3 four component-separated E-mode polarization maps analysed with the inpainting procedure: Sevem (red), SMICA (blue), NILC (green), and Commander (orange). The black dot corresponds to the direction measured using the Sevem PR4 polarization data. For reference, we also show the CMB dipole direction, and the directions observed in the temperature data for Sevem PR4 at $N_{side} = 64$ (red star) and $N_{side} = 2048$ (black star).}
    \label{Summary_Dir}
\end{figure}

\section{Discussion and conclusions} \label{Con}

We have applied the local variance estimator to the latest Planck release data set (PR4) in both temperature and polarization, performing a frequentist analysis of the isotropy assumption. In temperature, we have not found any simulation from a set of 600 with an amplitude as large as the one observed in the data for $4^{\circ}$, $6^{\circ}$, and $8^{\circ}$ discs. This means that the data exhibits a dipolar-like behaviour across the sky at a level below 0.17\% for that range of angular scales. We have performed the analysis at two different resolutions, $N_{side} = 64$ and $N_{side} = 2048$. For low resolution maps the direction points towards (l, b) = ($208^{\circ}$, $-15^{\circ}$), while for full resolution the direction is (l, b) = ($205^{\circ}$, $-20^{\circ}$). These results are in good agreement with previous reports, and can not be explained by residual systematics since a similar detection is found in WMAP.

Regarding polarization, we have implemented an alternative inpainting approach in order to minimize some effects that affect the reconstruction of the E-mode map from the measured Stokes parameters. This technique works optimally under certain conditions as will be shown in a future work (Gimeno-Amo et al. in preparation) that are not completely satisfied in the Planck data. The main problem is the low number of realistic noise simulations, which does not allow one to compute in an accurate way the noise covariance matrix. However, we have checked in this work that the inpainting is not biasing the parameters of the analysis and that also improves significantly the results compared with a simple masking approach.

First of all, we have applied the inpainting technique to the PR3 data set getting results that are in good agreement with the ones obtained by the Planck collaboration using a different inpainting algorithm. 
The small differences found can be explained by the large intrinsic dispersion of the estimated dipole direction, as well as differences in the considered mask and inpainting algorithm.
In particular, we tested that both the p-values and the directions are affected by the considered mask, with the exception of Sevem,
whose p-value is always below 1\%.

For PR4, we have performed the analysis only for Sevem. We get a p-value that lies between 1\%-3\% for sky fractions in the range from 67\% to 59.4\%. For smaller $f_{\rm sky}$ the p-value increases rapidly, getting up to 7.4\% for $f_{\rm sky}$ = 57\%. If the assymetry is of cosmological origin, this behaviour is somehow expected due to the loss of information when the number of available pixels is reduced. 
However, we can not discard that this effect is related to the reduction of foreground contamination in the CMB map when increasing the mask, that one would expect if residuals were the cause of the observed assymetry. Finally, the results are probably even more sensitive to the sky fraction and geometry of the mask, due to the fact that the estimated direction of the dipole is close to its boundary.

Regarding the differences between both pipelines, 
we tested that the values obtained for PR3 (0.22\%) and PR4 (2.8\%) are fully consistent taking into account the differences between both data sets (number of simulations, mask, inpainting, noise properties and transfer function).

It is also interesting to note that the amplitude of the dipole estimated for intensity ($\sim$7\%) and polarization ($\sim$9\%) are of similar level and point in a similar direction. If the intensity dipole were simply a statistical fluke such that the standard model still holds, we could expect a similar dipolar pattern in the E-mode map, taking into account the correlation between T and E. 
However, since this correlation is small (typically around 10 per cent), a much lower amplitude in the dipole in polarization would be expected and, therefore, this seems inconsistent with the values found. Even more, we have checked with E-mode maps constrained to intensity modulated simulations, that this correlation is expected to increase the p-value of the alignment by only about $0.5\%$ when realistic E2E noise is included. In any case, given the large uncertainties associated to the estimated amplitudes, improved polarization data is needed before any further conclusion can be established.

We may wonder whether the difference found between the p-values of PR3 and PR4 (0.22\% for PR3 versus 2.8\% for PR4 for Sevem for our reference case) is significant. There are several effects that can cause this difference. First of all, especially since we are in the tail of the distribution, the limited number of simulations may play a role. Indeed, if we consider the 95 per cent range for the p-values of both pipelines, they overlap, even if this range only takes into account the dispersion due to the different possible splits. Additional uncertainty would come from the estimator itself, including the inpainting. Another point to take into account is that we are using a different mask. Although we fix the same threshold in the maximum error of the reconstruction of the E-mode, this leads to a different sky fraction for PR3 and PR4, which can also influence the found p-value. Nevertheless, there are other reasons related to the data themselves that could lead to these results. In particular, if the anomaly were due to systematic effects, it would be expected to find a lower significance for PR4, due to the improvement in this regard in the data. However, if the anomaly were of cosmological origin, the significance could also be reduced due to the presence of the transfer function that affects the low multipoles in polarization for PR4. Therefore, no clear conclusion can be derived from these results, except that a modest detection of asymmetry is still present in the E-mode polarization maps.

Unfortunately, the sensitivity of the data is still not sufficiently good to obtain robust results. In particular, from the model-dependent test done in this paper, we see how the model parameters present large uncertainties. Furthermore, with the Planck polarization data set would be only possible to claim a detection with a p-value lower than 1\% in 95\% of the cases if the amplitude of the modulation were at the level of 16\%, well above the values found. Therefore, significant improvement in the sensitivity and systematics of the polarization data is needed on large angular scales before a robust result can be achieved in relation to possible violation of isotropy. Future experiments, such as LiteBIRD, that will be able to measure the cosmological E-modes at the cosmic-variance limit over a very large fraction of the sky, are expected to provide key information on the HPA.

\acknowledgments

We thank Yashar Akrami, Patricio Vielva, Patricia Diego-Palazuelos, Matteo Billi, Reijo Keskitalo, and Elena de la Hoz for valueble advice and comments. 
The authors would like to thank the Spanish Agencia Estatal
de Investigación (AEI, MICIU) for the financial support provided
under the project PID2019-110610RB-C21, as well as support from Universidad de Cantabria and Consejer\'{\i}a de Universidades, Igualdad, Cultura y Deporte from the Gobierno de Cantabria through the \emph{Instrumentaci\'on y ciencia de datos para sondear la naturaleza del universo} project.
CGA also thanks the funding from the Formación de Personal Investigador (FPI, Ref. PRE2020-096429) program of the Spanish Ministerio de Ciencia, Innovaci\'on y Universidades. 
The presented results are based on observations obtained with Planck\footnote{\url{http://www.esa.int/Planck}}, an ESA
science mission with instruments and contributions directly funded
by ESA Member States, NASA, and Canada. This research used resources of the National Energy Research Scientific Computing Center (NERSC), a U.S. Department of Energy Office of Science User Facility located at Lawrence Berkeley National Laboratory, operated under Contract No. DE-AC02-05CH11231. The results of this paper have been derived using the HEALPix package \cite{Gorski:2004by}, and the healpy \cite{Zonca2019}, numpy \cite{2020Natur.585..357H}, matplotlib \cite{2007CSE.....9...90H}, scipy \cite{2020NatMe..17..261V}, and dask\footnote{\url{https://dask.org}} Python packages.








\bibliographystyle{JHEP}
\bibliography{Bibliography}

\providecommand{\href}[2]{#2}\begingroup\raggedright\begin{thebibliography}{10}

\bibitem{2020A&A...641A..10P}
{\scshape Planck} collaboration, \emph{{Planck 2018 results. X. Constraints on inflation}}, \href{https://doi.org/10.1051/0004-6361/201833887}{\emph{Astron. Astrophys.} {\bfseries 641} (2020) A10} [\href{https://arxiv.org/abs/1807.06211}{{\ttfamily 1807.06211}}].

\bibitem{Planck:2018nkj}
{\scshape Planck} collaboration, \emph{{Planck 2018 results. I. Overview and the cosmological legacy of Planck}}, \href{https://doi.org/10.1051/0004-6361/201833880}{\emph{Astron. Astrophys.} {\bfseries 641} (2020) A1} [\href{https://arxiv.org/abs/1807.06205}{{\ttfamily 1807.06205}}].

\bibitem{2020A&A...641A...5P}
{\scshape Planck} collaboration, \emph{{Planck 2018 results. V. CMB power spectra and likelihoods}}, \href{https://doi.org/10.1051/0004-6361/201936386}{\emph{Astron. Astrophys.} {\bfseries 641} (2020) A5} [\href{https://arxiv.org/abs/1907.12875}{{\ttfamily 1907.12875}}].

\bibitem{2020A&A...641A...6P}
{\scshape Planck} collaboration, \emph{{Planck 2018 results. VI. Cosmological parameters}}, \href{https://doi.org/10.1051/0004-6361/201833910}{\emph{Astron. Astrophys.} {\bfseries 641} (2020) A6} [\href{https://arxiv.org/abs/1807.06209}{{\ttfamily 1807.06209}}].

\bibitem{Planck:2019evm}
{\scshape Planck} collaboration, \emph{{Planck 2018 results. VII. Isotropy and Statistics of the CMB}}, \href{https://doi.org/10.1051/0004-6361/201935201}{\emph{Astron. Astrophys.} {\bfseries 641} (2020) A7} [\href{https://arxiv.org/abs/1906.02552}{{\ttfamily 1906.02552}}].

\bibitem{2003ApJS..148....1B}
C.~Bennett et~al., \emph{{First-Year Wilkinson Microwave Anisotropy Probe (WMAP) Observations: Preliminary Maps and Basic Results}}, \href{https://doi.org/10.1086/377253}{\emph{Astrophys. J. Suppl.} {\bfseries 148} (2003) 1} [\href{https://arxiv.org/abs/astro-ph/0302207}{{\ttfamily astro-ph/0302207}}].

\bibitem{2007ApJS..170..288H}
G.~Hinshaw et~al., \emph{{Three-Year Wilkinson Microwave Anisotropy Probe (WMAP) Observations: Temperature Analysis}}, \href{https://doi.org/10.1086/513698}{\emph{Astrophys. J. Suppl.} {\bfseries 170} (2007) 288} [\href{https://arxiv.org/abs/astro-ph/0603451}{{\ttfamily astro-ph/0603451}}].

\bibitem{2009ApJS..180..225H}
G.~Hinshaw et~al., \emph{{Five-Year Wilkinson Microwave Anisotropy Probe Observations: Data Processing, Sky Maps, and Basic Results}}, \href{https://doi.org/10.1088/0067-0049/180/2/225}{\emph{Astrophys. J. Suppl.} {\bfseries 180} (2009) 225} [\href{https://arxiv.org/abs/0803.0732}{{\ttfamily 0803.0732}}].

\bibitem{2011ApJS..192...17B}
C.~Bennett et~al., \emph{{Seven-year Wilkinson Microwave Anisotropy Probe (WMAP) Observations: Are There Cosmic Microwave Background Anomalies?}}, \href{https://doi.org/10.1088/0067-0049/192/2/17}{\emph{Astrophys. J. Suppl.} {\bfseries 192} (2011) 17} [\href{https://arxiv.org/abs/1001.4758}{{\ttfamily 1001.4758}}].

\bibitem{2013ApJS..208...19H}
G.~Hinshaw et~al., \emph{{Nine-year Wilkinson Microwave Anisotropy Probe (WMAP) Observations: Cosmological Parameter Results}}, \href{https://doi.org/10.1088/0067-0049/208/2/19}{\emph{Astrophys. J. Suppl.} {\bfseries 208} (2013) 19} [\href{https://arxiv.org/abs/1212.5226}{{\ttfamily 1212.5226}}].

\bibitem{2004ApJ...605...14E}
H.K.~{Eriksen}, F.K.~{Hansen}, A.J.~{Banday}, K.M.~{G{\'o}rski} and P.B.~{Lilje}, \emph{{Asymmetries in the Cosmic Microwave Background Anisotropy Field}}, \href{https://doi.org/10.1086/382267}{\emph{Astrophys. J.} {\bfseries 605} (2004) 14} [\href{https://arxiv.org/abs/astro-ph/0307507}{{\ttfamily astro-ph/0307507}}].

\bibitem{2004MNRAS.354..641H}
F.K.~{Hansen}, A.J.~{Banday} and K.M.~{G{\'o}rski}, \emph{{Testing the cosmological principle of isotropy: local power-spectrum estimates of the WMAP data}}, \href{https://doi.org/10.1111/j.1365-2966.2004.08229.x}{\emph{MNRAS} {\bfseries 354} (2004) 641} [\href{https://arxiv.org/abs/astro-ph/0404206}{{\ttfamily astro-ph/0404206}}].

\bibitem{2004PhRvD..69f3516D}
A.~{de Oliveira-Costa}, M.~{Tegmark}, M.~{Zaldarriaga} and A.~{Hamilton}, \emph{{Significance of the largest scale CMB fluctuations in WMAP}}, \href{https://doi.org/10.1103/PhysRevD.69.063516}{\emph{Phys. Rev. D} {\bfseries 69} (2004) 063516} [\href{https://arxiv.org/abs/astro-ph/0307282}{{\ttfamily astro-ph/0307282}}].

\bibitem{2004ApJ...609...22V}
P.~{Vielva}, E.~{Mart{\'\i}nez-Gonz{\'a}lez}, R.B.~{Barreiro}, J.L.~{Sanz} and L.~{Cay{\'o}n}, \emph{{Detection of Non-Gaussianity in the Wilkinson Microwave Anisotropy Probe First-Year Data Using Spherical Wavelets}}, \href{https://doi.org/10.1086/421007}{\emph{Astrophys. J.} {\bfseries 609} (2004) 22} [\href{https://arxiv.org/abs/astro-ph/0310273}{{\ttfamily astro-ph/0310273}}].

\bibitem{2010AdAst2010E..77V}
P.~{Vielva}, \emph{{A Comprehensive Overview of the Cold Spot}}, \href{https://doi.org/10.1155/2010/592094}{\emph{Advances in Astronomy} {\bfseries 2010} (2010) 592094} [\href{https://arxiv.org/abs/1008.3051}{{\ttfamily 1008.3051}}].

\bibitem{Monteserin08}
C.~{Monteser{\'\i}n}, R.B.~{Barreiro}, P.~{Vielva}, E.~{Mart{\'\i}nez-Gonz{\'a}lez}, M.P.~{Hobson} and A.N.~{Lasenby}, \emph{{A low cosmic microwave background variance in the Wilkinson Microwave Anisotropy Probe data}}, \href{https://doi.org/10.1111/j.1365-2966.2008.13149.x}{\emph{MNRAS} {\bfseries 387} (2008) 209} [\href{https://arxiv.org/abs/0706.4289}{{\ttfamily 0706.4289}}].

\bibitem{2006A&A...454..409B}
A.~{Bernui}, T.~{Villela}, C.A.~{Wuensche}, R.~{Leonardi} and I.~{Ferreira}, \emph{{On the cosmic microwave background large-scale angular correlations}}, \href{https://doi.org/10.1051/0004-6361:20054243}{\emph{Astron. Astrophys.} {\bfseries 454} (2006) 409} [\href{https://arxiv.org/abs/astro-ph/0601593}{{\ttfamily astro-ph/0601593}}].

\bibitem{2023CQGra..40i4001K}
P.~{K. Aluri}, P.~{Cea}, P.~{Chingangbam}, M.-C.~{Chu}, R.G.~{Clowes}, D.~{Hutsem{\'e}kers} et~al., \emph{{Is the observable Universe consistent with the cosmological principle?}}, \href{https://doi.org/10.1088/1361-6382/acbefc}{\emph{Classical and Quantum Gravity} {\bfseries 40} (2023) 094001} [\href{https://arxiv.org/abs/2207.05765}{{\ttfamily 2207.05765}}].

\bibitem{2007ApJ...660L..81E}
H.K.~{Eriksen}, A.J.~{Banday}, K.M.~{G{\'o}rski}, F.K.~{Hansen} and P.B.~{Lilje}, \emph{{Hemispherical Power Asymmetry in the Third-Year Wilkinson Microwave Anisotropy Probe Sky Maps}}, \href{https://doi.org/10.1086/518091}{\emph{Astrophys. J. Lett.} {\bfseries 660} (2007) L81} [\href{https://arxiv.org/abs/astro-ph/0701089}{{\ttfamily astro-ph/0701089}}].

\bibitem{2009ApJ...704.1448H}
F.K.~{Hansen}, A.J.~{Banday}, K.M.~{G{\'o}rski}, H.K.~{Eriksen} and P.B.~{Lilje}, \emph{{Power Asymmetry in Cosmic Microwave Background Fluctuations from Full Sky to Sub-Degree Scales: Is the Universe Isotropic?}}, \href{https://doi.org/10.1088/0004-637X/704/2/1448}{\emph{Astrophys. J.} {\bfseries 704} (2009) 1448} [\href{https://arxiv.org/abs/0812.3795}{{\ttfamily 0812.3795}}].

\bibitem{2009ApJ...699..985H}
J.~{Hoftuft}, H.K.~{Eriksen}, A.J.~{Banday}, K.M.~{G{\'o}rski}, F.K.~{Hansen} and P.B.~{Lilje}, \emph{{Increasing Evidence for Hemispherical Power Asymmetry in the Five-Year WMAP Data}}, \href{https://doi.org/10.1088/0004-637X/699/2/985}{\emph{Astrophys. J.} {\bfseries 699} (2009) 985} [\href{https://arxiv.org/abs/0903.1229}{{\ttfamily 0903.1229}}].

\bibitem{2014A&A...571A..23P}
{\scshape Planck} collaboration, \emph{{Planck 2013 results. XXIII. Isotropy and statistics of the CMB}}, \href{https://doi.org/10.1051/0004-6361/201321534}{\emph{Astron. Astrophys.} {\bfseries 571} (2014) A23} [\href{https://arxiv.org/abs/1303.5083}{{\ttfamily 1303.5083}}].

\bibitem{Akrami:2014eta}
Y.~Akrami, Y.~Fantaye, A.~Shafieloo, H.K.~Eriksen, F.K.~Hansen, A.J.~Banday et~al., \emph{{Power asymmetry in WMAP and Planck temperature sky maps as measured by a local variance estimator}}, \href{https://doi.org/10.1088/2041-8205/784/2/L42}{\emph{Astrophys. J. Lett.} {\bfseries 784} (2014) L42} [\href{https://arxiv.org/abs/1402.0870}{{\ttfamily 1402.0870}}].

\bibitem{2015MNRAS.446.4232A}
S.~{Adhikari}, \emph{{Local variance asymmetries in Planck temperature anisotropy maps}}, \href{https://doi.org/10.1093/mnras/stu2408}{\emph{MNRAS} {\bfseries 446} (2015) 4232} [\href{https://arxiv.org/abs/1408.5396}{{\ttfamily 1408.5396}}].

\bibitem{2016A&A...594A..16P}
{\scshape Planck} collaboration, \emph{{Planck 2015 results. XVI. Isotropy and statistics of the CMB}}, \href{https://doi.org/10.1051/0004-6361/201526681}{\emph{Astron. Astrophys.} {\bfseries 594} (2016) A16} [\href{https://arxiv.org/abs/1506.07135}{{\ttfamily 1506.07135}}].

\bibitem{2005PhRvD..72j3002G}
C.~{Gordon}, W.~{Hu}, D.~{Huterer} and T.~{Crawford}, \emph{{Spontaneous isotropy breaking: A mechanism for CMB multipole alignments}}, \href{https://doi.org/10.1103/PhysRevD.72.103002}{\emph{Phys. Rev. D} {\bfseries 72} (2005) 103002} [\href{https://arxiv.org/abs/astro-ph/0509301}{{\ttfamily astro-ph/0509301}}].

\bibitem{2007ApJ...656..636G}
C.~{Gordon}, \emph{{Broken Isotropy from a Linear Modulation of the Primordial Perturbations}}, \href{https://doi.org/10.1086/510511}{\emph{Astrophys. J.} {\bfseries 656} (2007) 636} [\href{https://arxiv.org/abs/astro-ph/0607423}{{\ttfamily astro-ph/0607423}}].

\bibitem{2017PhRvD..95f3011Z}
J.P.~{Zibin} and D.~{Contreras}, \emph{{Testing physical models for dipolar asymmetry: From temperature to k space to lensing}}, \href{https://doi.org/10.1103/PhysRevD.95.063011}{\emph{Phys. Rev. D} {\bfseries 95} (2017) 063011} [\href{https://arxiv.org/abs/1512.02618}{{\ttfamily 1512.02618}}].

\bibitem{2015PhRvD..91b3515R}
P.K.~{Rath}, P.K.~{Aluri} and P.~{Jain}, \emph{{Relating the inhomogeneous power spectrum to the CMB hemispherical anisotropy}}, \href{https://doi.org/10.1103/PhysRevD.91.023515}{\emph{Phys. Rev. D} {\bfseries 91} (2015) 023515} [\href{https://arxiv.org/abs/1403.2567}{{\ttfamily 1403.2567}}].

\bibitem{2009PhRvD..80f3004H}
D.~{Hanson} and A.~{Lewis}, \emph{{Estimators for CMB statistical anisotropy}}, \href{https://doi.org/10.1103/PhysRevD.80.063004}{\emph{Phys. Rev. D} {\bfseries 80} (2009) 063004} [\href{https://arxiv.org/abs/0908.0963}{{\ttfamily 0908.0963}}].

\bibitem{2016JCAP...01..046G}
S.~{Ghosh}, R.~{Kothari}, P.~{Jain} and P.K.~{Rath}, \emph{{Dipole modulation of cosmic microwave background temperature and polarization}}, \href{https://doi.org/10.1088/1475-7516/2016/01/046}{\emph{JCAP} {\bfseries 2016} (2016) 046} [\href{https://arxiv.org/abs/1507.04078}{{\ttfamily 1507.04078}}].

\bibitem{2023arXiv230104539D}
Dipanshu, T.~Souradeep and S.~Hirve, \emph{{Capturing Statistical Isotropy Violation with Generalized Isotropic Angular Correlation Functions of Cosmic Microwave Background Anisotropy}}, \href{https://doi.org/10.3847/1538-4357/ace895}{\emph{Astrophys. J.} {\bfseries 954} (2023) 181} [\href{https://arxiv.org/abs/2301.04539}{{\ttfamily 2301.04539}}].

\bibitem{Marcos-Caballero:2019jqj}
A.~Marcos-Caballero and E.~Mart\'\i{}nez-Gonz\'alez, \emph{{Scale-dependent dipolar modulation and the quadrupole-octopole alignment in the CMB temperature}}, \href{https://doi.org/10.1088/1475-7516/2019/10/053}{\emph{JCAP} {\bfseries 10} (2019) 053} [\href{https://arxiv.org/abs/1909.06093}{{\ttfamily 1909.06093}}].

\bibitem{2008PhRvD..78l3520E}
A.L.~{Erickcek}, M.~{Kamionkowski} and S.M.~{Carroll}, \emph{{A hemispherical power asymmetry from inflation}}, \href{https://doi.org/10.1103/PhysRevD.78.123520}{\emph{Phys. Rev. D} {\bfseries 78} (2008) 123520} [\href{https://arxiv.org/abs/0806.0377}{{\ttfamily 0806.0377}}].

\bibitem{2009PhRvD..80b3526D}
J.F.~{Donoghue}, K.~{Dutta} and A.~{Ross}, \emph{{Nonisotropy in the CMB power spectrum in single field inflation}}, \href{https://doi.org/10.1103/PhysRevD.80.023526}{\emph{Phys. Rev. D} {\bfseries 80} (2009) 023526} [\href{https://arxiv.org/abs/astro-ph/0703455}{{\ttfamily astro-ph/0703455}}].

\bibitem{2010PhRvD..81h3501C}
S.M.~{Carroll}, C.-Y.~{Tseng} and M.B.~{Wise}, \emph{{Translational invariance and the anisotropy of the cosmic microwave background}}, \href{https://doi.org/10.1103/PhysRevD.81.083501}{\emph{Phys. Rev. D} {\bfseries 81} (2010) 083501} [\href{https://arxiv.org/abs/0811.1086}{{\ttfamily 0811.1086}}].

\bibitem{2011JHEP...02..061K}
T.S.~{Koivisto} and D.F.~{Mota}, \emph{{CMB statistics in noncommutative inflation}}, \href{https://doi.org/10.1007/JHEP02(2011)061}{\emph{Journal of High Energy Physics} {\bfseries 2011} (2011) 61} [\href{https://arxiv.org/abs/1011.2126}{{\ttfamily 1011.2126}}].

\bibitem{2013JCAP...08..007L}
D.H.~{Lyth}, \emph{{The CMB modulation from inflation}}, \href{https://doi.org/10.1088/1475-7516/2013/08/007}{\emph{JCAP} {\bfseries 2013} (2013) 007} [\href{https://arxiv.org/abs/1304.1270}{{\ttfamily 1304.1270}}].

\bibitem{2016MNRAS.460.1577K}
R.~{Kothari}, S.~{Ghosh}, P.K.~{Rath}, G.~{Kashyap} and P.~{Jain}, \emph{{Imprint of inhomogeneous and anisotropic primordial power spectrum on CMB polarization}}, \href{https://doi.org/10.1093/mnras/stw1039}{\emph{MNRAS} {\bfseries 460} (2016) 1577} [\href{https://arxiv.org/abs/1503.08997}{{\ttfamily 1503.08997}}].

\bibitem{2022arXiv220903928S}
K.~{Sravan Kumar} and J.~{Marto}, \emph{{Hemispherical asymmetry of primordial power spectra}}, \href{https://doi.org/10.48550/arXiv.2209.03928}{\emph{arXiv e-prints} (2022) arXiv:2209.03928} [\href{https://arxiv.org/abs/2209.03928}{{\ttfamily 2209.03928}}].

\bibitem{Hansen:2023gra}
F.K.~Hansen, E.F.~Boero, H.E.~Luparello and D.G.~Lambas, \emph{{A possible common explanation for several cosmic microwave background (CMB) anomalies: A strong impact of nearby galaxies on observed large-scale CMB fluctuations}}, \href{https://doi.org/10.1051/0004-6361/202346779}{\emph{Astron. Astrophys.} {\bfseries 675} (2023) L7} [\href{https://arxiv.org/abs/2305.00268}{{\ttfamily 2305.00268}}].

\bibitem{2012MNRAS.420.2162F}
R.~{Fern{\'a}ndez-Cobos}, P.~{Vielva}, R.B.~{Barreiro} and E.~{Mart{\'\i}nez-Gonz{\'a}lez}, \emph{{Multiresolution internal template cleaning: an application to the Wilkinson Microwave Anisotropy Probe 7-yr polarization data}}, \href{https://doi.org/10.1111/j.1365-2966.2011.20182.x}{\emph{MNRAS} {\bfseries 420} (2012) 2162} [\href{https://arxiv.org/abs/1106.2016}{{\ttfamily 1106.2016}}].

\bibitem{2008ApJ...676...10E}
H.K.~{Eriksen}, J.B.~{Jewell}, C.~{Dickinson}, A.J.~{Banday}, K.M.~{G{\'o}rski} and C.R.~{Lawrence}, \emph{{Joint Bayesian Component Separation and CMB Power Spectrum Estimation}}, \href{https://doi.org/10.1086/525277}{\emph{Astrophys. J.} {\bfseries 676} (2008) 10} [\href{https://arxiv.org/abs/0709.1058}{{\ttfamily 0709.1058}}].

\bibitem{2020MNRAS.492.3994G}
S.~{Ghosh} and P.~{Jain}, \emph{{A pixel space method for testing dipole modulation in the CMB polarization}}, \href{https://doi.org/10.1093/mnras/stz3627}{\emph{MNRAS} {\bfseries 492} (2020) 3994} [\href{https://arxiv.org/abs/1807.02359}{{\ttfamily 1807.02359}}].

\bibitem{2017arXiv171000580A}
P.K.~{Aluri} and A.~{Shafieloo}, \emph{{Power asymmetry in CMB polarization maps from PLANCK : a local variance analysis}}, \href{https://doi.org/10.48550/arXiv.1710.00580}{\emph{arXiv e-prints} (2017) arXiv:1710.00580} [\href{https://arxiv.org/abs/1710.00580}{{\ttfamily 1710.00580}}].

\bibitem{Planck:2020olo}
{\scshape Planck} collaboration, \emph{{Planck intermediate results. LVII. Joint Planck LFI and HFI data processing}}, \href{https://doi.org/10.1051/0004-6361/202038073}{\emph{Astron. Astrophys.} {\bfseries 643} (2020) A42} [\href{https://arxiv.org/abs/2007.04997}{{\ttfamily 2007.04997}}].

\bibitem{Gorski:2004by}
K.M.~{G{\'o}rski}, E.~{Hivon}, A.J.~{Banday}, B.D.~{Wandelt}, F.K.~{Hansen}, M.~{Reinecke} et~al., \emph{{HEALPix: A Framework for High-Resolution Discretization and Fast Analysis of Data Distributed on the Sphere}}, \href{https://doi.org/10.1086/427976}{\emph{Astrophys. J.} {\bfseries 622} (2005) 759} [\href{https://arxiv.org/abs/astro-ph/0409513}{{\ttfamily astro-ph/0409513}}].

\bibitem{Planck:2018yye}
{\scshape Planck} collaboration, \emph{{Planck 2018 results. IV. Diffuse component separation}}, \href{https://doi.org/10.1051/0004-6361/201833881}{\emph{Astron. Astrophys.} {\bfseries 641} (2020) A4} [\href{https://arxiv.org/abs/1807.06208}{{\ttfamily 1807.06208}}].

\bibitem{2008arXiv0803.1814C}
J.-F.~{Cardoso}, M.~{Martin}, J.~{Delabrouille}, M.~{Betoule} and G.~{Patanchon}, \emph{{Component separation with flexible models. Application to the separation of astrophysical emissions}}, \href{https://doi.org/10.48550/arXiv.0803.1814}{\emph{arXiv e-prints} (2008) arXiv:0803.1814} [\href{https://arxiv.org/abs/0803.1814}{{\ttfamily 0803.1814}}].

\bibitem{10.1111/j.1365-2966.2011.19770.x}
S.~Basak and J.~Delabrouille, \emph{{A needlet internal linear combination analysis of WMAP 7-year data: estimation of CMB temperature map and power spectrum}}, \href{https://doi.org/10.1111/j.1365-2966.2011.19770.x}{\emph{Monthly Notices of the Royal Astronomical Society} {\bfseries 419} (2011) 1163}.

\bibitem{Zonca2019}
A.~Zonca, L.~Singer, D.~Lenz, M.~Reinecke, C.~Rosset, E.~Hivon et~al., \emph{healpy: equal area pixelization and spherical harmonics transforms for data on the sphere in python}, \href{https://doi.org/10.21105/joss.01298}{\emph{Journal of Open Source Software} {\bfseries 4} (2019) 1298}.

\bibitem{Tegmark:2001zv}
M.~Tegmark and A.~de~Oliveira-Costa, \emph{{How to measure CMB polarization power spectra without losing information}}, \href{https://doi.org/10.1103/PhysRevD.64.063001}{\emph{Phys. Rev. D} {\bfseries 64} (2001) 063001} [\href{https://arxiv.org/abs/astro-ph/0012120}{{\ttfamily astro-ph/0012120}}].

\bibitem{Lewis:2003an}
A.~Lewis, \emph{{Harmonic E/B decomposition for CMB polarization maps}}, \href{https://doi.org/10.1103/PhysRevD.68.083509}{\emph{Phys. Rev. D} {\bfseries 68} (2003) 083509} [\href{https://arxiv.org/abs/astro-ph/0305545}{{\ttfamily astro-ph/0305545}}].

\bibitem{bar08}
R.B.~{Barreiro}, P.~{Vielva}, C.~{Hernandez-Monteagudo} and E.~{Martinez-Gonzalez}, \emph{{A Linear Filter to Reconstruct the ISW Effect From CMB and LSS Observations}}, \href{https://doi.org/10.1109/JSTSP.2008.2005350}{\emph{IEEE Journal of Selected Topics in Signal Processing} {\bfseries 2} (2008) 747} [\href{https://arxiv.org/abs/0809.2557}{{\ttfamily 0809.2557}}].

\bibitem{Kothari:2018cmt}
R.~Kothari, \emph{{A comprehensive study of Modulation effects on CMB Polarization}}, \href{https://doi.org/10.1007/s10714-022-02921-8}{\emph{Gen. Rel. Grav.} {\bfseries 54} (2018) 2022} [\href{https://arxiv.org/abs/1806.06325}{{\ttfamily 1806.06325}}].

\bibitem{2020Natur.585..357H}
C.R.~{Harris}, K.J.~{Millman}, S.J.~{van der Walt}, R.~{Gommers}, P.~{Virtanen}, D.~{Cournapeau} et~al., \emph{{Array programming with NumPy}}, \href{https://doi.org/10.1038/s41586-020-2649-2}{\emph{Nature} {\bfseries 585} (2020) 357} [\href{https://arxiv.org/abs/2006.10256}{{\ttfamily 2006.10256}}].

\bibitem{2007CSE.....9...90H}
J.D.~{Hunter}, \emph{{Matplotlib: A 2D Graphics Environment}}, \href{https://doi.org/10.1109/MCSE.2007.55}{\emph{Computing in Science and Engineering} {\bfseries 9} (2007) 90}.

\bibitem{2020NatMe..17..261V}
P.~{Virtanen}, R.~{Gommers}, T.E.~{Oliphant}, M.~{Haberland}, T.~{Reddy}, D.~{Cournapeau} et~al., \emph{{SciPy 1.0: fundamental algorithms for scientific computing in Python}}, \href{https://doi.org/10.1038/s41592-019-0686-2}{\emph{Nature Methods} {\bfseries 17} (2020) 261} [\href{https://arxiv.org/abs/1907.10121}{{\ttfamily 1907.10121}}].

\end{thebibliography}\endgroup

\end{document}